\documentclass[prd,12pt]{article}

\usepackage{tikz}
\usepackage{amsmath,amssymb,graphicx,multirow,xspace,slashed,array,booktabs}
\usepackage[colorlinks=true,urlcolor=blue,anchorcolor=blue,citecolor=blue,filecolor=blue,linkcolor=blue,menucolor=blue,pagecolor=blue]{hyperref}
\usepackage[compress,numbers]{natbib}
\usepackage{subfigure, placeins}
\usepackage{booktabs}
\usepackage{blindtext, rotating}
\usepackage{afterpage}
\usepackage{enumitem}
\usepackage{ marvosym }
\usepackage{authblk} 
\usepackage{verbatim}
\usepackage{soul} 
\usepackage[normalem]{ulem}
\usepackage{pifont}

\usepackage[utf8]{inputenc}

\usepackage{floatrow}
\newfloatcommand{capbtabbox}{table}[][\FBwidth]

\usepackage[font=footnotesize,labelfont=bf]{caption}

\usepackage{lineno}

\usepackage{bm}

\allowdisplaybreaks

\long\def\symbolfootnote[#1]#2{\begingroup%
\def\thefootnote{\fnsymbol{footnote}}\footnote[#1]{#2}\endgroup}

\newcommand{\newc}{\newcommand}
\newc{\gsim}{\lower.7ex\hbox{$\;\stackrel{\textstyle>}{\sim}\;$}}
\newc{\lsim}{\lower.7ex\hbox{$\;\stackrel{\textstyle<}{\sim}\;$}}
\newc{\gev}{\,{\rm GeV}}
\newc{\mev}{\,{\rm MeV}}
\newc{\ev}{\,{\rm eV}}
\newc{\kev}{\,{\rm keV}}
\newc{\tev}{\,{\rm TeV}}
\newc{\MHT}{$H_T^{\text{miss}}$}
\newc{\MET}{$\slashed{E}_T$}
\newc{\MTT}{$M_{T2}$}

\newc{\mz}{M_Z}
\newc{\mpl}{M_*}
\newc{\mw}{m_{\rm weak}}
\newc{\nr}[1]{N^c_R{}_{#1}}

\definecolor{darkgreen}{rgb}{0,0.5,0}
\definecolor{goodorange}{rgb}{0.9,0.4,0}

\def\beq{\begin{equation}}
\def\eeq{\end{equation}}
\newcommand{\bea}{\begin{eqnarray}\begin{aligned}}
\newcommand{\eea}{\end{aligned}\end{eqnarray}}
\def\bitem{\begin{itemize}}
\def\eitem{\end{itemize}}

\numberwithin{equation}{section}

\newcommand\fverb{\setbox\fverbbox=\hbox\bgroup\verb}

\newbox\fverbbox

\newcommand\CSRL{{C^S_{RL}}}
\newcommand\CSLL{{C^S_{LL}}}
\newcommand\CVLL{{C^V_{LL}}}
\newcommand\CVRL{{C^V_{RL}}}
\newcommand\CTLL{{C^T_{LL}}}

\definecolor{darkturquoise}{rgb}{0.0, 0.81, 0.82}
\definecolor{amaranth}{rgb}{0.9, 0.17, 0.31}
\definecolor{awesome}{rgb}{1.0, 0.13, 0.32}
\definecolor{ballblue}{rgb}{0.13, 0.67, 0.8}
\definecolor{blue(munsell)}{rgb}{0.0, 0.5, 0.69}
\definecolor{cerulean}{rgb}{0.0, 0.48, 0.65}
\definecolor{darkcyan}{rgb}{0.0, 0.55, 0.55}
\definecolor{darklavender}{rgb}{0.45, 0.31, 0.59}
\definecolor{darkmagenta}{rgb}{0.55, 0.0, 0.55}
\definecolor{deepfuchsia}{rgb}{0.76, 0.33, 0.76}
\definecolor{emerald}{rgb}{0.31, 0.78, 0.47}
\definecolor{goldenpoppy}{rgb}{0.99, 0.76, 0.0}
\definecolor{green(pigment)}{rgb}{0.0, 0.65, 0.31}
\definecolor{jazzberryjam}{rgb}{0.65, 0.04, 0.37}
\definecolor{jade}{rgb}{0.0, 0.66, 0.42}
\definecolor{princetonorange}{rgb}{1.0, 0.56, 0.0}
\definecolor{richelectricblue}{rgb}{0.03, 0.57, 0.82}

\begin{document}

\baselineskip 0.6 cm

\begin{titlepage}

\thispagestyle{empty}

		\begin{flushright}
			MIT-CTP/5215\\
			P3H-20-027 
		\end{flushright}

\begin{center}

\vskip 1cm

{\LARGE \bf A Complete Framework for }\vskip0.3cm
{\LARGE \bf Tau Polarimetry in $B\to D^{(\ast)}\tau\nu$ Decays}

\vskip 0.5cm

\vskip 1cm
{\large Pouya Asadi$^1$, Anna Hallin$^2$, Jorge Martin Camalich$^{3,4}$,}\\
\vskip 0.2cm
{\large David Shih$^{2,5,6}$, Susanne Westhoff$^7$}
\end{center}

\vskip 0.5 cm

\noindent {\it  $^1$ Center for Theoretical Physics, Massachusetts Institute of Technology\\
\phantom{C}Cambridge, MA 02139, USA} \\
{\it  $^2$ NHETC, Dept.~of Physics and Astronomy, Rutgers University\\
\phantom{C}Piscataway, NJ 08854 USA}\\
{\it  $^3$ Instituto de Astrof\'isica de Canarias, C/ V\'ia L\'actea, s/n
E38205\\
\phantom{C}La Laguna, Tenerife, Spain}\\
{\it  $^4$ Universidad de La Laguna, Departamento de Astrof\'isica, La Laguna, Tenerife, Spain}\\
{\it  $^5$ Physics Division, Lawrence Berkeley National Laboratory, Berkeley, CA 94720, USA} \\
{\it  $^6$ Center for Theoretical Physics, University of California, Berkeley, CA 94720, USA}\\
{\it  $^7$ Institute for Theoretical Physics, Heidelberg University, 69120 Heidelberg, Germany} \\
\vskip 0.5cm

\abstract{\noindent The meson decays $B\to D\tau\nu$ and $B\to D^\ast \tau \nu$ are sensitive probes of the $b\to c\tau\nu$ transition. In this work we present a complete framework to obtain the maximum information on the
physics of $B\to D^{(\ast)}\tau\nu$ with polarized $\tau$ leptons and unpolarized $D^{(\ast)}$ mesons.
Focusing on the hadronic decays $\tau\to \pi\nu$ and $\tau\to\rho\nu$, we show how to extract seven $\tau$ asymmetries from a fully differential analysis of the final-state kinematics. At Belle II with $50~\text{ab}^{-1}$ of data, these asymmetries could potentially be measured with percent level statistical uncertainty. This would open a new window into possible new physics contributions in $b\to c\tau\nu$ and would allow us to decipher its Lorentz and gauge structure.}

\flushbottom

\end{titlepage}

\tableofcontents

\newpage

\setcounter{page}{1}

\section{Introduction}\label{sec:intro}

\noindent Leptonic and semileptonic hadron decays are important probes of the fundamental quark-lepton interactions within and beyond the Standard Model (SM). Decays of $B$ mesons with $\tau$ leptons in the final state, in particular, provide a unique way to determine the properties of fermion interactions involving the third generation. They allow us to test the flavor structure of the SM and search for New Physics (NP) predominantly coupled to the heavier fermions. In addition, the large $\tau$ mass leads to an enhanced sensitivity to the scalar component of the weak interaction. Semitauonic $B$ decays are therefore especially sensitive to the time-like component of the virtual $W$ boson~\cite{Korner:1989qb} or to the exchange of new (pseudo)scalar particles~\cite{Tanaka:1994ay,Kiers:1997zt,Goldberger:1999yh}.

At flavor experiments, the decays $B\to D\tau\nu$ and $B\to D^\ast\tau\nu$, both triggered by the charged-current transition $b\to c\tau\nu$, are the most accessible semitauonic hadron decays. The branching ratios of these decays normalized to those into light leptons, $R_{D^{(*)}}={\rm BR}(B\to D^{(*)}\tau\nu)/{\rm BR}(B\to D^{(*)}\ell\nu)$ with $\ell=e,\,\mu$, have been measured with good precision at BaBar~\cite{Lees:2013uzd,Lees:2012xj}, Belle~\cite{Huschle:2015rga,Sato:2016svk,Abdesselam:2019dgh} and LHCb~\cite{Aaij:2017uff,Aaij:2017deq}. Interestingly, the combination of these measurements appears to be about $20\%$ larger than the SM prediction with a significance of $3.08\,\sigma$~\cite{hflav}. The normalized branching fraction of $B_c\to J/\psi \tau\nu$, which is based on the same $b\to c\tau\nu$ transition, has been measured by LHCb and also appears to be larger than the SM expectation~\cite{Aaij:2017tyk}.
Beyond total rates, in $B\to D^\ast\tau\nu$ the longitudinal $\tau$ polarization $P_L(\tau)$~\cite{Abdesselam:2016xqt,Hirose:2017dxl} and the fraction of longitudinally polarized $D^\ast$ mesons $F_L(D^\ast)$~\cite{Abdesselam:2019wbt,Adamczyk:2019wyt} have been measured. This shows the potential of the current flavor experiments, Belle II and LHCb, to extract the properties of the $b\to c\tau\nu$ transition by measuring the $\tau$ kinematics in the decay. Precise analyses of these transitions are important to understand the origin of the observed discrepancies with the SM and to decipher the structure of NP in case they persist.

Due to its fast decay, the production properties of the $\tau$ lepton cannot be directly measured, but have to be extracted from the decay products where part of the information on the $\tau$ momentum is carried away by at least one neutrino in the final state. Extracting the properties of the $b\to c\tau\nu$ transition from the visible $\tau$ decay products in $B\to D^{(\ast)}\tau\nu$ has evolved into a comprehensive research program~\cite{Bullock:1992yt,Lange:2001uf,Nierste:2008qe,Tanaka:2010se,Hagiwara:2014tsa,Bordone:2016tex,Alonso:2016gym,Ligeti:2016npd,Alonso:2017ktd,Asadi:2018sym,Alonso:2018vwa,Bernlochner:2020tfi,Bhattacharya:2020lfm}. One aims to construct the full differential decay rate and then integrate out all kinematic variables that are unobservable due to the presence of neutrinos~\cite{Alonso:2016gym,Bhattacharya:2020lfm}. 

In this paper, we develop a complete framework to extract the full set of $B\to D^{(\ast)}\tau\nu$ observables (with polarized $\tau$ and unpolarized $D^{(\ast)}$) from the visible final state. We focus on the hadronic $\tau$ decays $\tau\to \pi\nu$ and $\tau\to\rho\nu$, which preserve more information on the $\tau$ kinematics than the leptonic decays $\tau\to\ell\nu\nu$ \cite{Alonso:2017ktd}. In the two-body decays the $\tau$ spin orientation is directly imprinted on the pion or rho direction of flight. The $\tau$ helicity and kinematics can thus be deduced from the energy and angular distributions of the visible final-state particles~\cite{Tanaka:2010se,Sakaki:2012ft,Alonso:2017ktd,Ivanov:2017mrj}.  The main result of our paper allows us to express the differential decay rate of $B\to D^{(\ast)}\tau(\to d\nu)\nu$ as
\beq\label{eq:intro_alt}
\frac{d^3\Gamma_d}{dq^2 d\cos\theta_d ds_d}=n(q^2)\left(1+ \sum_{{\mathcal O}}F_{\mathcal O}^d(q^2,\cos\theta_d,s_d){\mathcal O}(q^2)\right).
\eeq
Here $q^2$, $\cos\theta_d$ and $s_d$ describe measurable kinematic quantities (the momentum transfer to the lepton pair; the angle between the $\tau$ daughter $d$ and the $D^{(\ast)}$; the energy of the $d$) in the leptonic rest frame, and $n(q^2)$ is a normalization factor. Importantly, the ``leptonic functions" $F_{\mathcal O}^d(q^2,\cos\theta_d,s_d)$ depend only on the $\tau\to d\nu$ decay. The sum is over seven
 asymmetry observables of the $B\to D^{(\ast)}\tau\nu$ transition,
\beq\label{eq:intro_observables}
{\mathcal O}=A_{FB},\,P_L,\,P_\perp,\,Z_L,\,Z_\perp,\,Z_Q,\,A_Q,
\eeq
to be defined in the next section. This formula directly relates these asymmetry observables to the kinematic distribution of the $\tau$ daughter. By measuring the kinematics of the $d$ particle, one can extract nearly all the physics of the $b\to c\tau\nu$ transition, including the possible presence of new physics affecting the transition. 

The asymmetry observables represent a useful intermediate step between the data and the underlying Wilson coefficients. Previous studies~\cite{Tanaka:1994ay,Sakaki:2012ft,Ivanov:2017mrj,Fajfer:2012vx,Tanaka:2012nw,Datta:2012qk,Duraisamy:2013pia,Biancofiore:2013ki,Duraisamy:2014sna,Ivanov:2015tru,Bhattacharya:2015ida,Alok:2016qyh,Becirevic:2016hea,Bardhan:2016uhr,Chen:2017eby,Azatov:2018knx,Dai:2018eom,Becirevic:2019tpx,Alguero:2020ukk,Shi:2019gxi} have identified a subset of these asymmetries ($A_{FB}$, $P_L$, $P_\perp$) and shown how to extract them from differential distributions of the final state. In this work,
we show that a total of nine asymmetries, together with the differential decay rate $d\Gamma_B/ dq^2$, suffice to describe the full physics of $B\to D^{(*)}\tau\nu$ with unpolarized $D^{(*)}$. The remaining two not listed in (\ref{eq:intro_observables}), $P_T$ and $Z_T$, are nonzero only in the presence of CP violation, and furthermore are only accessible by including additional information, e.g.\ from $D^{(\ast)}$ decays~\cite{Bhattacharya:2020lfm}, in the kinematic distributions. We reserve a complete study of these additional observables for a future publication~\cite{future}.

We will demonstrate how one could theoretically measure the asymmetries by performing an unbinned maximum likelihood fit to the $d$ distribution (\ref{eq:intro_alt}). While we do not include realistic experimental considerations such as systematic uncertainties, detector acceptance or backgrounds (these are beyond the scope of this work), we show that at least the statistical power  with 50~ab$^{-1}$ of Belle II data should be enough to measure the asymmetry observables to  percent level precision. 

Analytic formulas like (\ref{eq:intro_alt})  could prove useful in experimental studies. Besides being needed for maximum likelihood fits, they could be adapted for Monte Carlo generators \cite{Aebischer:2019zoe}. Another line in this direction
has been providing efficient methods to reweight Monte Carlo event samples interpreting experimental data directly in terms of SM or NP parameters~\cite{Lange:2001uf,Ligeti:2016npd,Bernlochner:2020tfi}.

The outline of the paper is as follows. In Section~\ref{sec:obsB} we decompose the $B\rightarrow D^{(*)}\tau\nu$ kinematics into a complete set of $\tau$ asymmetries. These asymmetries contain all information that could be obtained if the $\tau$ momentum was fully accessible. In Section~\ref{sec:fulldecay} we show how to  extract seven of the nine $\tau$ asymmetries from the kinematics of the $\tau$ decay products. By performing a full-fledged statistical analysis in Section~\ref{sec:stats}, we give a theoretical estimate of the expected sensitivity of Belle II to the asymmetries, assuming a given number of events and neglecting experimental effects. We also demonstrate how to decipher the structure of new physics in $\tau$ production in the framework of an effective theory and in context of the current anomalies found in the $R_{D^{(\ast)}}$ ratios. We conclude in Section~\ref{sec:discuss} with a summary and outlook.

\section{Tau asymmetries in $B\rightarrow D^{(\ast)}\tau\nu$}
\label{sec:obsB}

\noindent In this section we focus on the $B\rightarrow M\tau\nu$ decay kinematics, where $M=D$ or $D^*$, without considering the $\tau$ decays yet. The narrow width of the $\tau$ enables a factorization of the full decay chain into a $\tau$ production part and a $\tau$ decay part.

The basis for the $\tau$ asymmetries is the differential decay rate for $B\rightarrow M\tau\nu$ with the $\tau$ spin quantized along an arbitrary direction $\mathbf{\hat{e}}_a$,
\begin{equation}
\label{dGammaBfundamental}
	d\Gamma^{\lambda_\tau,a}_{B}=\frac{1}{2m_B}\big|\mathcal{M}_{B}^{\lambda_\tau,a}\big|^2d\Phi_3(p_B;p_{M},p_{\tau},p_{\nu})\,.
\end{equation}
Here $\lambda_\tau=\pm$ is the direction of the $\tau$ spin along the $\mathbf{\hat{e}}_a$ axis, and the Lorentz invariant phase space for a particle $i$ decaying to $n$ daughters is
\begin{equation}
	d\Phi_n(p_i;p_1,...,p_n)=(2\pi)^4 \prod_{j=1}^n \frac{d^3\mathbf{p}_j}{(2\pi)^3 2 E_j}\,\delta^4\!\!\left(p_i-\sum_{j=1}^n p_j\right) .
\end{equation}
Throughout this work we sum over the polarization states of the $D^*$ meson. 

The total differential decay rate can be calculated from the spin-dependent decay rates along any axis as
\begin{equation}
	d\Gamma_{B}=d\Gamma^{+,a}_{B}+d\Gamma^{-,a}_{B}\,.
\end{equation}
On the contrary, a $\tau$ spin asymmetry
\begin{equation}
\label{asymmeqn}
	d\mathcal{P}^{a}_{B}=d\Gamma^{+,a}_{B}-d\Gamma^{-,a}_{B}
\end{equation}
is always defined along the particular axis $\mathbf{\hat{e}}_{a}$. 

We work in the ``$q^2$ frame'', the center of mass frame of the lepton pair, with $q^2=(p_B-p_M)^2$ being the momentum squared transferred to the leptons. Fig.~\ref{fig:angles} illustrates the various momentum vectors, polarization vectors and angles involved in the $q^2$ frame. Let $\{\mathbf{\hat{e}}_1,\mathbf{\hat{e}}_2,\mathbf{\hat{e}}_3\}$ be an orthonormal coordinate system in this frame and choose
\begin{equation}
\label{e3}
		\mathbf{\hat{e}}_3=\mathbf{\hat{p}}_{\tau}\equiv\mathbf{\hat{e}}_L\,, 
\end{equation}
where $\mathbf{\hat{p}}_{\tau}$ is the direction of the $\tau$ momentum. The spin-dependent differential decay rate \eqref{dGammaBfundamental} and the asymmetries \eqref{asymmeqn} can then be expressed using $\tau$ helicity amplitudes
$\mathcal{M}_{B}^{\lambda_{\tau},L}$. From here on, the index $a=L$ will be suppressed. The resulting expressions for the asymmetries depend on how the axes $\hat{\mathbf{e}}_1$ and $\hat{\mathbf{e}}_2$ are chosen. Choosing
\begin{equation}
\label{e2e1}
	\mathbf{\hat{e}}_2=\frac{\mathbf{\hat{p}}_{M}\times\mathbf{\hat{p}}_{\tau}}{|\mathbf{\hat{p}}_{M}\times\mathbf{\hat{p}}_{\tau}|}\equiv\mathbf{\hat{e}}_T\,, \quad \mathbf{\hat{e}}_1=\mathbf{\hat{e}}_T\times\mathbf{\hat{e}}_L \equiv\mathbf{\hat{e}}_{\perp}\,,
\end{equation}
where $\mathbf{\hat{p}}_{M}$ is the direction of the $M$ momentum, results in
\begin{equation}
\begin{split}
\label{GammaB}
	d\Gamma^{\lambda_{\tau}}_{B}&=\frac{1}{2m_B}\big|\mathcal{M}_{B}^{\lambda_{\tau}}\big|^2d\Phi_3(p_B;p_{M},p_{\tau},p_{\nu})\,,\\
	d\mathcal{P}_{B}^{\perp}&=\frac{1}{2m_B}2\,\mathrm{Re}\left[\mathcal{M}_{B}^{+}\big(\mathcal{M}_{B}^{-}\big)^{\dagger}\right]d\Phi_3(p_B;p_{M},p_{\tau},p_{\nu})\,,\\
		d\mathcal{P}_{B}^{T}&=\frac{1}{2m_B}2\,\mathrm{Im}\left[\mathcal{M}_{B}^{+}\big(\mathcal{M}_{B}^{-}\big)^{\dagger}\right]d\Phi_3(p_B;p_{M},p_{\tau},p_{\nu})\,.
\end{split}
\end{equation}
These four differential distributions capture all the information in the matrix elements $\mathcal{M}_B^\pm$ in the $B \rightarrow D^{(*)}\tau\nu$ decay (with unpolarized $D^*$).

The matrix elements do not depend on the azimuthal angle of the $\tau$ momentum with respect to the $M$ momentum, this angle is thus integrated out. The two remaining degrees of freedom in the final state are chosen to be $q^2$ and $\cos\theta_{\tau}$, where $\theta_{\tau}$ is the angle between the flight direction of the $\tau$ and the negative
direction of the $M$ momentum in the $q^2$ frame.  The decay rates and asymmetries for $B\rightarrow M\tau\nu$ can be expanded in spherical harmonics encoding conservation of angular momentum~\cite{Korner:1989qb},
\bea\label{cosexp}
\frac{d^2\Gamma_{B}^{\lambda_{\tau}}}{dq^2 d\cos\theta_{\tau}}&={d\Gamma_B\over dq^2}\sum_{\ell=0}^2
B^{\lambda_\tau}_\ell(q^2) P_\ell^0(\cos\theta_\tau)\cr
\frac{d^2\mathcal{P}_{B}^{\perp}}{dq^2 d\cos\theta_{\tau}} &=
{d\Gamma_B\over dq^2}\sum_{\ell=1}^2 {\rm Re}[C_\ell(q^2)]P_\ell^1(\cos\theta_\tau)\cr
 \frac{d^2\mathcal{P}_{B}^T}{dq^2 d\cos\theta_{\tau}} &=
{d\Gamma_B\over dq^2}\sum_{\ell=1}^2 {\rm Im}[C_\ell(q^2)]P_\ell^1(\cos\theta_\tau)\,.
\eea
where $P_\ell^{0,1}(\cos\theta_\tau)$ are the associated Legendre functions. 
Together with the total differential rate $d\Gamma_B/dq^2$, the angular coefficient functions $B^\pm_{0,1,2}(q^2)\in\mathbb{R}$ and $C_{1,2}(q^2)\in\mathbb{C}$ 
describe the full kinematic information in $B\to M\tau\nu
$ decays with unpolarized mesons.

Although the angular coefficient functions are themselves perfectly valid observables, it is more conventional (and physical) to work in terms of various asymmetries of the $\tau$ angle $\cos\theta_\tau$ and spin direction $\lambda_\tau$. An equivalent and complete basis of $\tau$ asymmetries is as follows:

\begin{itemize}
\item The $\tau$ forward-backward asymmetry is 
\beq\label{eq:dAFBdq}
A_{FB}(q^2) = \left({d\Gamma_B\over dq^2}\right)^{-1}\left(\int_{0}^1 d\cos\theta_\tau
- \int_{-1}^0 d\cos\theta_\tau\right){d^2\Gamma_B\over dq^2d\cos\theta_\tau} = B_1^+ + B_1^-\,.
\eeq

\item Pure spin asymmetries are given by:
\beq\label{eq:dPdq}
P_a(q^2) =  \left({d\Gamma_B\over dq^2}\right)^{-1}\int_{-1}^1 d\cos\theta_\tau\,\, {d^2\mathcal{P}_B^a\over dq^2d\cos\theta_\tau} =\begin{cases} 2(B_0^+ - B_0^-) & a=L\cr
-{\pi\over2}\mathrm{Re}[C_1] & a=\perp\cr
-{\pi\over2}\mathrm{Im}[C_1] & a=T\,.
\end{cases}
\eeq
The spin asymmetries $P_a$ correspond to the net longitudinal, perpendicular and transverse polarizations of the $\tau$ in $B\to M\tau\nu$ decays. The asymmetries we have described so far have been considered before in the literature~\cite{Tanaka:1994ay,Sakaki:2012ft,Ivanov:2017mrj,Fajfer:2012vx,Tanaka:2012nw,Datta:2012qk,Duraisamy:2013pia,Biancofiore:2013ki,Duraisamy:2014sna,Ivanov:2015tru,Bhattacharya:2015ida,Alok:2016qyh,Becirevic:2016hea,Bardhan:2016uhr,Chen:2017eby,Azatov:2018knx,Dai:2018eom,Becirevic:2019tpx,Alguero:2020ukk,Shi:2019gxi}.

\item We can additionally consider {\it double asymmetries} with respect to both $\lambda_\tau$ and $\cos\theta_\tau$,\footnote{$Z$ stands for {\it zweifach}.}
\beq\label{eq:dAdq}
Z_a(q^2)= \left({d\Gamma_B\over dq^2} \right)^{-1} \left(\int_{0}^1 d\cos\theta_\tau
- \int_{-1}^0 d\cos\theta_\tau\right){d^2\mathcal{P}_B^a\over dq^2d\cos\theta_\tau}=\begin{cases} B_1^+-B_1^- & a=L\cr
-2\mathrm{Re}[C_2] & a=\perp\cr
-2\mathrm{Im}[C_2] & a=T\,.\end{cases}
\eeq
These asymmetries have not been considered before and give access to the previously unexplored combinations of angular coefficient functions $B_1^\pm$ and $C_2$.\footnote{In $B\to D\tau\nu$ some of the $\tau$ asymmetries are absent because the pseudoscalar nature of the $D$ meson restricts the possible angular coefficient functions. In particular, it has been shown that $B_1^-=0$ for the most general effective Lagrangian with scalar, vector and tensor operators (with left-handed neutrinos only)~\cite{Asadi:2018sym}. This implies that $Z_L=A_{FB}$ in $B\to D\tau\nu$ and no independent information is gained from $Z_L$.
In $B\to D^\ast\tau\nu$, $B_1^-$ is generated by the transverse polarization of the $D^\ast$ vector meson \cite{Alonso:2016gym}, so that $A_{FB}$ and $Z_L$ carry independent information.}

\item The angular coefficient functions $B_2^\pm$ cannot be expressed in terms of simple asymmetries like the other functions. They denote the \textit{quadrupole} part of the partial wave expansion in (\ref{GammaB}). We combine these angular coefficient functions to define the following asymmetry observables
\begin{eqnarray}
A_{Q}(q^2) &=& \left({d\Gamma_B\over dq^2}\right)^{-1} \frac{5}{2} \int_{-1}^1 d\cos\theta_\tau\,  P_2^0 (\cos \theta_\tau)  {d^2\Gamma_B\over dq^2d\cos\theta_\tau}  = B_2^+ + B_2^-,  \\
Z_{Q}(q^2) &=& \left({d\Gamma_B\over dq^2}\right)^{-1} \frac{5}{2} \int_{-1}^1 d\cos\theta_\tau\,  P_2^0 (\cos \theta_\tau) {d^2\mathcal{P}_B^L\over dq^2d\cos\theta_\tau}  = B_2^+ - B_2^-, \nonumber
\label{eq:dAQdq}
\end{eqnarray}
where the $5/2$ prefactor captures the Legendre polynomial normalization.

\item Finally, given that we have extracted an overall factor of $d\Gamma_B/dq^2$ in the definition (\ref{cosexp}) of the angular coefficient functions, they satisfy the relation $B_0^++B_0^-={1\over2}$.

\end{itemize}

\section{Tau asymmetries from the visible final state}
\label{sec:fulldecay}

Since the $\tau$ decays promptly in the detector with one or two neutrinos in the final state, it is generally not possible to reconstruct its full four-momentum.\footnote{The full $\tau$ kinematics could potentially be accessed with displaced 3-prong decays~\cite{Hill:2019zja}.} Therefore the $\tau$ asymmetries in $B\to M\tau\nu$ described in Section~\ref{sec:obsB} are not directly measurable. In this section we will show how they can be extracted from final-state observables with fully reconstructed mesons $M$. We focus on the two-body decays $\tau\to d\nu$ with $d=\pi,\rho$, as they preserve more information compared to the three-body decay $\tau\to\ell\nu\nu$. However, our formalism can be straightforwardly generalized to $\tau\to\ell\nu\nu$ or other $\tau$ decay modes. 

\begin{figure}[!t]
\centering
\includegraphics[scale=0.5]{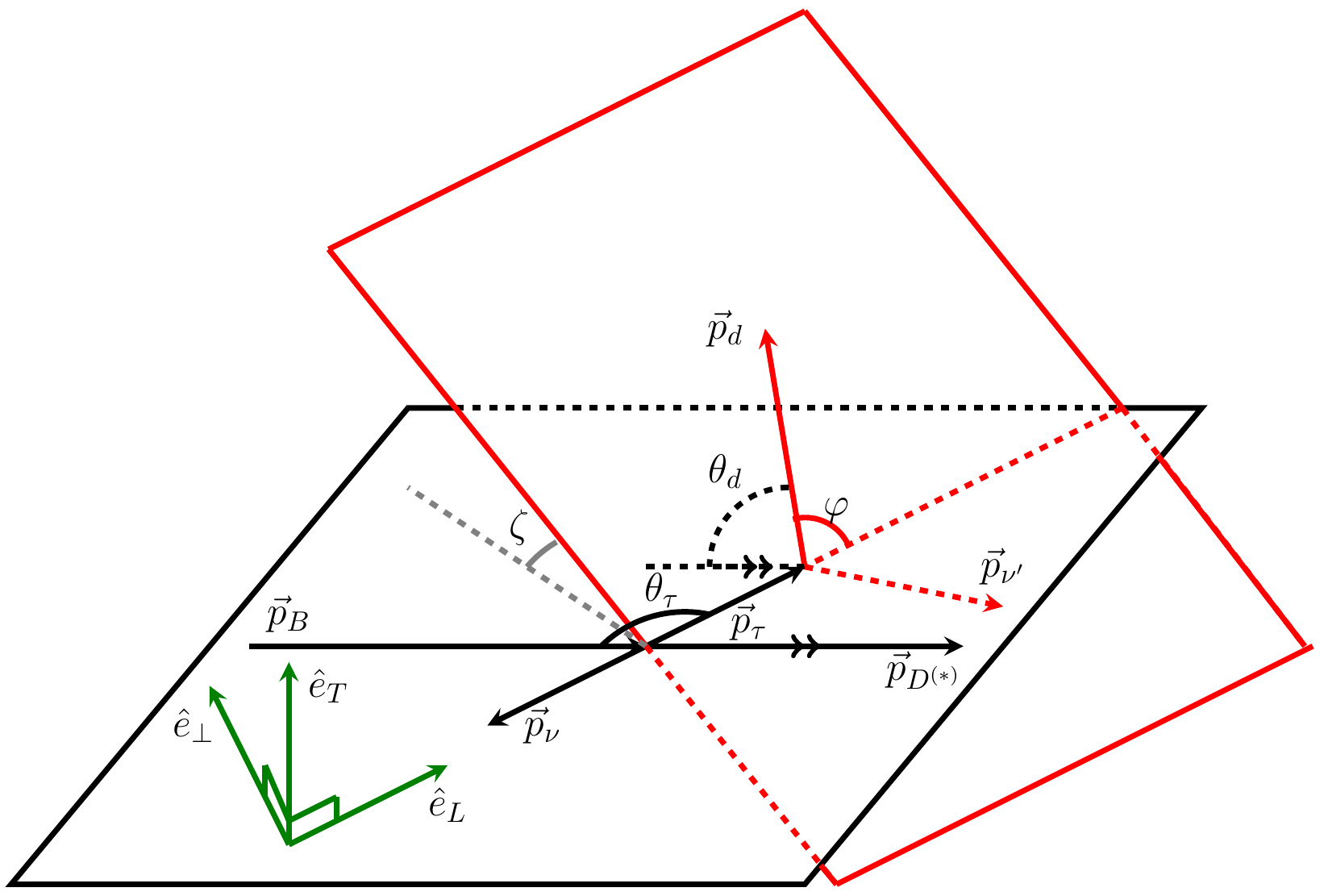}
\caption{The two decay planes of $B\rightarrow M\tau\nu$ (black) and $\tau\rightarrow d\nu$ (red) in the $q^2$ frame. The angle between the two planes is denoted $\zeta$. The angle between the flight direction of the $\tau$ and the negative direction of $M$ is denoted $\theta_{\tau}$. In the decay plane of the $\tau$, $\varphi$ is the angle between the direction of the $\tau$ and the direction of the daughter particle $d$. Finally, the angle $\theta_d$ is the angle between the direction of the daughter particle (in the $\tau$ decay plane) and the negative direction of $M$ (in the $B$ decay plane). }
\label{fig:angles}
\end{figure}

Fig.~\ref{fig:angles} shows the two decay planes of $B\rightarrow M\tau\nu$ and $\tau\rightarrow d\nu$, and the various angles and momenta involved in the decays in the $q^2$ frame. The angle $\theta_d$ between $\vec p_d$ and $-\vec p_{M}$ is the only directly measurable angle. Meanwhile $\varphi$ (the angle between $\vec p_\tau$ and $\vec p_d$), $\zeta$ (the angle between the two decay planes) and $\theta_\tau$ are not. However, $\varphi$ is completely determined by the kinematics of the two-body decay as
\begin{equation}
\label{cosvarphi}
\begin{split}
	\cos\varphi=\frac{(1+r_{\tau}^2)s_{d}-(r_{\tau}^2+r_{d}^2)}{(1-r_{\tau}^2)\sqrt{s_d^2-r_d^2}}\,.
\end{split}
\end{equation}
Here we have introduced the notation 
\beq\label{eq:notationsd}
s_d\equiv {E_d\over\sqrt{q^2}},\qquad r_{\tau}\equiv \frac{m_{\tau}}{\sqrt{q^2}},\qquad r_d \equiv \frac{m_d}{\sqrt{q^2}} \,,
\eeq
where $E_d$ and $m_d$ denote the energy and mass of the daughter particle in the $q^2$ frame. It is straightforward to show that the allowed range of $s_d$ is
\begin{equation}
s_d \in \left[ \frac{r_\tau^2}{2} \left(1 + \frac{r_d^2}{r_\tau^4} \right) ,\, \frac{1}{2} \left(1 + r_d^2\right) \right] .
\label{eq:rangesd}
\end{equation}
Finally, the angles $\theta_\tau$ and $\zeta$ are related to $\varphi$ and $\theta_d$ via
\beq
\label{changeofvariables}
	\cos\theta_d=\cos\theta_{\tau}\cos\varphi+\sin\theta_{\tau}\sin\varphi\cos\zeta\,.
\eeq

The matrix element for the full $B\rightarrow D^{(*)}\tau(\rightarrow d\nu_{2})\nu_{1}$ decay is
\begin{equation}
\label{matrelement}
	\mathcal{M}_{\mathrm{tot}} = \frac{1}{p_{\tau}^2-m_{\tau}^2+i m_{\tau}\Gamma_{\tau}}\sum_{\lambda_{\tau}=\pm}\mathcal{M}_{B}^{\lambda_{\tau}}\mathcal{M}_{\tau}^{\lambda_{\tau}},
\end{equation}
where $\mathcal{M}_{B}$ is the matrix element for $B\to M\tau\nu$ as introduced in Section~\ref{sec:obsB}, $\Gamma_\tau$ is the total width of the $\tau$, and $\mathcal{M}_{\tau}$ is the matrix element for $\tau\rightarrow d\nu$. 
In the narrow width approximation, the four-body phase space factorizes as
\begin{align}
\label{BWphasespace}
&	\frac{1}{(p_{\tau}^2-m_{\tau}^2)^2+m_{\tau}^2\Gamma_{\tau}^2}\, d\Phi_4(p_B;p_{M},p_{\nu_{1}},p_d,p_{\nu_{2}})\\\nonumber
& \hspace*{1cm}	\longrightarrow\frac{1}{2m_{\tau}\Gamma_{\tau}}\,d\Phi_3(p_B;p_{M},p_{\tau},p_{\nu_{1}}) \,d\Phi_2(p_{\tau};p_d,p_{\nu_{2}})
\end{align}
with the $\tau$ set on-shell. 
Then the full four-fold differential decay rate is
\begin{equation}
\label{4folddiff}
\begin{split}
	\frac{d^4\Gamma_d}{dq^2d\cos\theta_{\tau}ds_d\, d\zeta} & =\frac{E_{\tau}}{\Gamma_\tau m_{\tau}}\Bigg(\frac{d^2\Gamma_B^{\lambda_{\tau}}}{dq^2d\cos\theta_{\tau}}\frac{d^2\Gamma^{\lambda_{\tau}}_{\tau}}{ds_d\, d\zeta} \\
	&\qquad\qquad\qquad + \frac{1}{2} \left[ \frac{d^2\mathcal{P}_{B}^{{\perp}}}{dq^2 d\cos\theta_{\tau}}\frac{d^2\mathcal{P}_{\tau}^{\perp}}{ds_d\, d\zeta}-\frac{d^2\mathcal{P}_{B}^{{T}}}{dq^2 d\cos\theta_{\tau}}\frac{d^2\mathcal{P}_{\tau}^{T}}{ds_d\, d\zeta} \right] \Bigg),
\end{split}
\end{equation}
where repeated $\lambda_\tau$ indices are summed over. 
All terms are factorized into a $B$ decay part (see \eqref{cosexp}) and a $\tau$ decay part defined as
\begin{equation}
\label{Gammataulambdatau}
\begin{split}
	d\Gamma^{\lambda_{\tau}}_{\tau}&=\frac{1}{2E_{\tau}}\big|\mathcal{M}_{\tau}^{\lambda_{\tau}}\big|^2d\Phi_2(p_{\tau};p_d,p_{\nu_{2}})\,,\\
	d\mathcal{P}_{\tau}^{\perp}&=\frac{1}{2E_{\tau}}2\,\mathrm{Re}[\mathcal{M}_{\tau}^{+}(\mathcal{M}_{\tau}^{-})^{\dagger}]d\Phi_2(p_{\tau};p_d,p_{\nu_{2}})\,,\\
	d\mathcal{P}_{\tau}^{T}&=\frac{1}{2E_{\tau}}2\,\mathrm{Im}[\mathcal{M}_{\tau}^+(\mathcal{M}_{\tau}^-)^{\dagger}]d\Phi_2(p_{\tau};p_d,p_{\nu_{2}})\,.
\end{split}
\end{equation}
Similarly to how $d\Gamma^{\lambda_{\tau}}_{B}$ and $d\mathcal{P}_{B}^{\perp,T}$ could be expanded in $\cos\theta_\tau$, the expressions above can be expanded in the angles $\varphi$ and $\zeta$.
For a two-body $\tau\to d\nu$ decay, 
\begin{equation}
\label{tauangles}
\begin{split}
	\frac{d\Gamma^{\lambda_{\tau}}_{\tau}}{ds_d d\zeta}&=\frac{m_\tau \Gamma_{\tau\to d\nu} }{\pi E_\tau}\,g^{\lambda_{\tau}}_d(q^2,s_d) \,,\\
	\frac{d\mathcal{P}_{\tau}^{\perp}}{ds_d d\zeta}&=\frac{2m_\tau\Gamma_{\tau\to d\nu} }{\pi E_{\tau}}\,h_d(q^2,s_d)\sin\varphi (q^2,s_d)\cos\zeta \,,\\
	\frac{d\mathcal{P}_{\tau}^{T}}{ds_d d\zeta}&=\frac{2m_\tau \Gamma_{\tau\to d\nu} }{\pi E_{\tau}}\,h_d(q^2,s_d)\sin\varphi(q^2,s_d)\sin\zeta \,,
\end{split}
\end{equation}
where the coefficient functions for $d=\pi,\rho$ are given by
\bea
\label{gandh}
    & g_\pi^\pm=\frac{1}{1-r_{\tau}^2}\left(1\pm \frac{4s_{\pi}-(1+r_\tau^2)}{(1-r_\tau^2) }\right)\cr
    &	g^\pm_\rho
	=\frac{r_{\tau}^2\Big((1-r_\tau^2)(r_{\tau}^2-r_{\rho}^2)\left(2r_{\rho}^2+r_{\tau}^2\right)	\pm \left(r_\tau^2-2r_\rho^2\right)\left(4s_{\rho} r_\tau^2-(1+r_\tau^2)(r_\tau^2+r_\rho^2)\right)\Big) }{(r_{\tau}^2-r_{\rho}^2)^2(2r_{\rho}^2+r_{\tau}^2) (1-r_{\tau}^2)^2}\cr
	& h_\pi=\frac{2}{1-r_{\tau}^2} \frac{s_\pi}{r_{\tau}}\cr
	&h_\rho=\frac{2}{1-r_{\tau}^2} \left(\frac{r_\tau^2-2r_\rho^2}{2r_\rho^2+r_\tau^2}\right)\sqrt{s_\rho^2-r_\rho^2}\frac{r_{\tau}^3}{(r_{\tau}^2-r_{\rho}^2)^2}\,.
\eea
Throughout our analysis we neglect the $\pi$ mass but not the $\rho$ mass. 

To transform (\ref{4folddiff}) into a fully observable, fully differential decay rate, we need to integrate over the two unobservable angles $\theta_\tau$ and $\zeta$ and replace them with the single observable angle $\theta_d$. Formally this can be accomplished using (\ref{changeofvariables}) to obtain
\bea
\frac{d^3\Gamma_d}{dq^2d\cos\theta_{d }ds_d} = &	\int_{-1}^1 d\cos\theta_\tau\,\,\int_{-\pi}^\pi d\zeta\,\, \frac{d^4\Gamma_d}{dq^2d\cos\theta_{\tau}ds_d\, d\zeta}\cr &\qquad\qquad\qquad\qquad \times \delta(\cos\theta_d - \cos\theta_{\tau}\cos\varphi-\sin\theta_{\tau}\sin\varphi\cos\zeta)\,.
\label{d4GammaTod3Gamma}
\eea
In Appendix \ref{sec:angint} we carry out these integrals explicitly. The result is given by
\bea\label{explicitd3Gammachangevar}
	&\frac{d^3\Gamma_d}{dq^2d\cos\theta_{d}ds_d} = {\rm BR}(\tau\to d\nu) {d\Gamma_B\over dq^2} \sum_{\ell=0}^2 P_\ell(\cos\theta_d)I_\ell(q^2,s_d),\cr
	& I_0  
	= \frac{1}{2} \left(f_0^d(q^2)+f_L^d(q^2,s_d) P_L(q^2)\right) \cr
	&I_1  
	= f_{A_{FB}}^d(q^2,s_d)A_{FB}(q^2)+f_\perp^d(q^2,s_d)P_\perp(q^2)+f_{Z_L}^d(q^2,s_d)Z_L(q^2) \cr
	&I_2 
	=f_{Z_\perp}^d(q^2,s_d) Z_\perp(q^2)+f_{Z_Q}^d(q^2,s_d)Z_Q(q^2) + f_{A_Q}^d(q^2,s_d)A_Q(q^2) .
	\eea
Here we have used (\ref{eq:dAFBdq}) - (\ref{eq:dAQdq}) to connect the differential distribution to the $\tau$ asymmetries, and we have defined the leptonic functions
\bea
\label{f0fl}
	f_0^d(q^2)&=g^+_d(q^2,s_d)+g^-_d(q^2,s_d)\\ 
	f_L^d(q^2,s_d)&= g^+_d(q^2,s_d)-g^-_d(q^2,s_d)\cr
	f_{\perp}^d(q^2,s_d)&= {4\over\pi}  \sin^2\varphi\,h_d(q^2,s_d)\\
	f_{A_{FB}}^d(q^2,s_d)& = \cos\varphi \,f_0^d(q^2)\\
	f_{Z_L}^d(q^2,s_d)& =  \cos\varphi \,f_L^d(q^2,s_d)\\
	f_{Z_\perp}^d(q^2,s_d)
	&= \frac{3\pi}{4} \cos \varphi f_{\perp}^d(q^2,s_d)\\
	f_{A_Q}^d(q^2,s_d)&= \frac{1}{2} (3\cos^2\varphi-1) f_0^d(q^2)  \, \\ 
	f_{Z_Q}^d(q^2,s_d)&=  \frac{1}{2} (3\cos^2\varphi-1) f_L^d(q^2)  \,. 
\eea
One can verify using~\eqref{gandh} that the first four leptonic functions in (\ref{f0fl}) agree precisely with those considered in~\cite{Alonso:2017ktd}. 

We see that the fully differential final-state decay rate breaks down into a linear combination of the asymmetries, or equivalently, of the angular coefficient functions $B_{0,1,2}^\pm$ and $\mathrm{Re}[C_{1,2}]$. We emphasize that (\ref{explicitd3Gammachangevar}) is completely general even in the presence of arbitrary heavy new physics altering the $b\to c\tau\nu$ transition. The leptonic functions are independent of the $b\to c \tau\nu$ transition and depend only on the $\tau$ decay mode. Therefore one could use (\ref{explicitd3Gammachangevar})  to directly extract the asymmetries from the data in a completely model independent way. 
We will investigate the theoretical sensitivity of such an approach in the next section.

\section{Sensitivity to asymmetry observables at Belle II}
\label{sec:stats}

Having derived analytic expressions for the fully differential final-state decay rate and related them to the $B\to M\tau\nu$ asymmetry observables, we now turn to a toy study of how the asymmetry observables could be measured in practice, and what precision one could hope to achieve. We cannot comment on the systematic uncertainties associated with our proposal at different experiments. A detailed simulation of backgrounds and detector effects is also beyond the scope of this work. We will limit ourselves to calculating the achievable statistical uncertainty; this should furnish a ``best-case scenario" for the sensitivity of any future measurement. 

For the analysis in this section, we will need explicit numerical formulas for all the asymmetries in terms of the dimension-6 effective Hamiltonian. The dependence of the asymmetries $P_{a}(q^2)$ ($a=L,\perp,T$) and $A_{FB}(q^2)$ on all the relevant dimension-6 operators has already been calculated \cite{Asadi:2018sym}. Following the notation of \cite{Sakaki:2013bfa,Asadi:2018sym}, we include the analytic expressions for the new asymmetries $A_Q(q^2)$ and $Z_a(q^2)$ ($a=L,\perp,T,Q$) in Appendix~\ref{appx:analytic}. 

\subsection{Maximum likelihood method}

The energy $s_d$ and the angle $\cos \theta_d$ of the daughter in $\tau\to d\nu$ decays are directly measurable at Belle~II. Using the fully-differential distribution \eqref{explicitd3Gammachangevar}, we apply the unbinned maximum likelihood method in $s_d$ and $\cos\theta_d$ to fit for the asymmetry observables in $q^2$ bins and determine the covariance matrices around the best fit values. We do not assume any templates for the $q^2$ dependence from the SM or otherwise; 
we consider a separate and independent measurement of the asymmetries in each $q^2$ bin.\footnote{
The statistical analysis outlined in this section expands on a previous analysis of $B\rightarrow D \tau \nu$~\cite{Alonso:2017ktd}. In the analysis of~\cite{Alonso:2017ktd}, instead of a fit to the complete distribution of events in $\cos \theta_d$, only two bins distinguished by $\mathrm{sgn} \left( \cos \theta_d\right)$ are considered. By fitting to the full distribution, we get access to the new observables $Z_\perp$, $Z_L$, $Z_Q$ and $A_Q$, and we also increase the sensitivity to the remaining asymmetries.}

Let ${\mathcal O}(q_i^2)$
for $\mathcal{O}=A_{FB}$, $P_L$, etc.\ be the parameters that we want to fit for in $q^2$ bin $i$. According to \eqref{explicitd3Gammachangevar}, the probability distribution of events in $q^2$ bin $i$ in terms of these parameters is given by 
\bea
\label{sdkdist}
& p_i(s_d,\cos \theta_d\,|\mathcal{O}(q_i^2)) = {1\over2}\Big(f_0^d(q_i^2) + 
 f_L^d(s_d,q_i^2) P_L(q_i^2)  \Big) P_0(\cos \theta_d)\\ & \qquad  +  \Big( f_{A_{FB}}^d(s_d,q_i^2) A_{FB}(q_i^2) + f_{\perp}^d(s_d,q_i^2) P_{\perp}(q_i^2)
 +   f_{Z_L}^d(s_d,q_i^2) Z_L(q_i^2)   \Big) P_1(\cos \theta_d) \\ & \qquad 
 + \Big( f_{Z_\perp}^d(s_d,q_i^2) Z_{\perp}(q_i^2) + f_{Z_Q}^d(s_d,q_i^2) Z_Q(q_i^2) 
 +   f_{A_Q}^d(s_d,q_i^2) A_Q(q_i^2)    \Big) P_2(\cos \theta_d).
\eea
We assume the event numbers in each $q^2$ bin are large enough that the asymptotic form of the maximum likelihood method can be used. Then the log-likelihood statistic to be maximized is
\beq\label{eq:loglikeihoodint}
L(\mathcal{O}(q_i^2)) = N f_i \int ds_d\, d\cos\theta_d\,\, p_i(s_d,\cos\theta_d|\hat{\mathcal{O}}(q_i^2)) \log p_i(s_d,\cos\theta_d|\mathcal{O}(q_i^2)),
\eeq
where $\hat{\mathcal{O}}(q_i^2)$ are the true
 values of the asymmetry observables, $N$ is the total number of events, and
\beq
f_i \equiv \Delta q^2\Gamma_B^{-1}{d\Gamma_B\over dq^2}(q_i^2)
\eeq
is the fraction of events in $q^2$ bin $i$ with bin width $\Delta q^2$.
The elements of the inverse covariance matrix for bin $i$ are given by
\beq
(\Sigma^i)^{-1}_{ab} = -\partial_{\mathcal{O}_a(q_i^2)}\partial_{\mathcal{O}_b(q_i^2)}L(\mathcal{O}(q_i^2))\Big|_{\mathcal{O}(q_i^2)=\hat{\mathcal{O}}(q_i^2)}.
\eeq
In the following, we report the sensitivity to the $q^2$-integrated asymmetries, defined by
\begin{equation}
    \mathcal{O}=\frac{1}{\Gamma_B}\int dq^2 \frac{d\Gamma_B}{dq^2} \mathcal{O}(q^2),
\label{integratedobs2}
\end{equation}
where $\mathcal{O} = A_{FB},P_L,$ etc.
 These integrated asymmetries provide us with a sensitivity estimate in the case of limited event statistics.
To combine the covariance matrices in each $q^2$ bin into a total covariance matrix for the integrated observables, we use the discretized form of (\ref{integratedobs2}),
\beq
 \mathcal{O} \approx \sum_i f_i\,  {\mathcal O}(q_i^2).
\eeq
The total covariance matrix is then
\beq
    \Sigma = \sum_i f_i^2\,\Sigma^i\,.
\eeq
In the following subsections, we will report values and make plots of the variances $\sigma_a^2$ (the diagonal elements of $\Sigma$) and the correlation coefficients
 $\rho_{ab} = \sigma_{ab}/(\sigma_a\sigma_b)$ (derived from the off-diagonal elements of $\Sigma$).

\subsection{Standard Model sensitivity}

Using this method we determine the theoretical sensitivity to the $q^2$-integrated asymmetries assuming the SM prediction for all the parameters, i.e.\ $\hat{\mathcal{O}}(q_i^2) = \mathcal{O}_{SM}(q_i^2)$. It is trivial to repeat the analysis for a scenario with a different prior.

\begin{figure}
\centering
\includegraphics[scale=0.75]{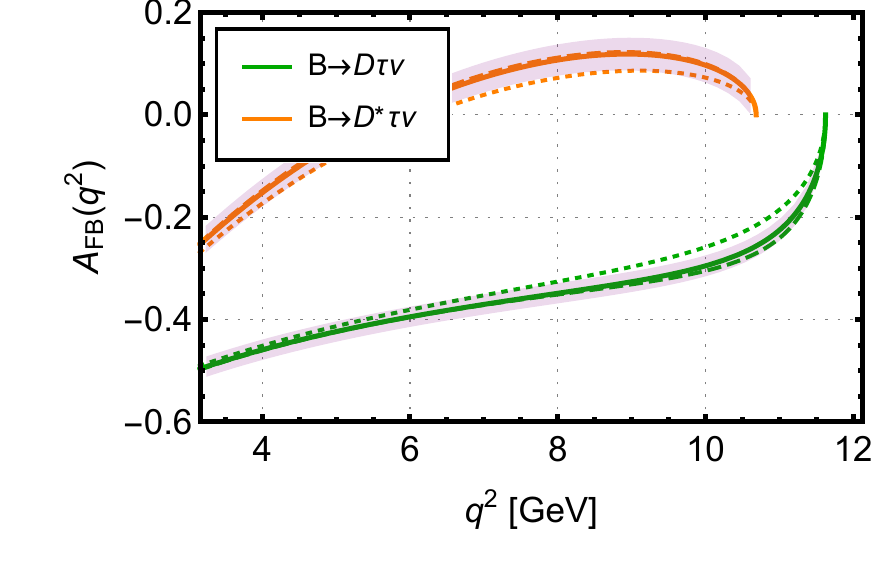} \\
\includegraphics[scale=0.75]{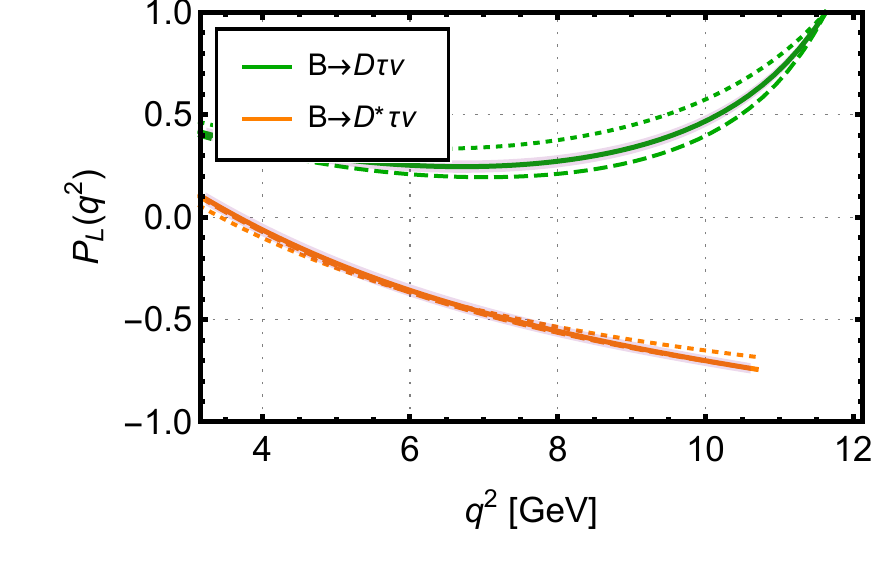} 
\includegraphics[scale=0.75]{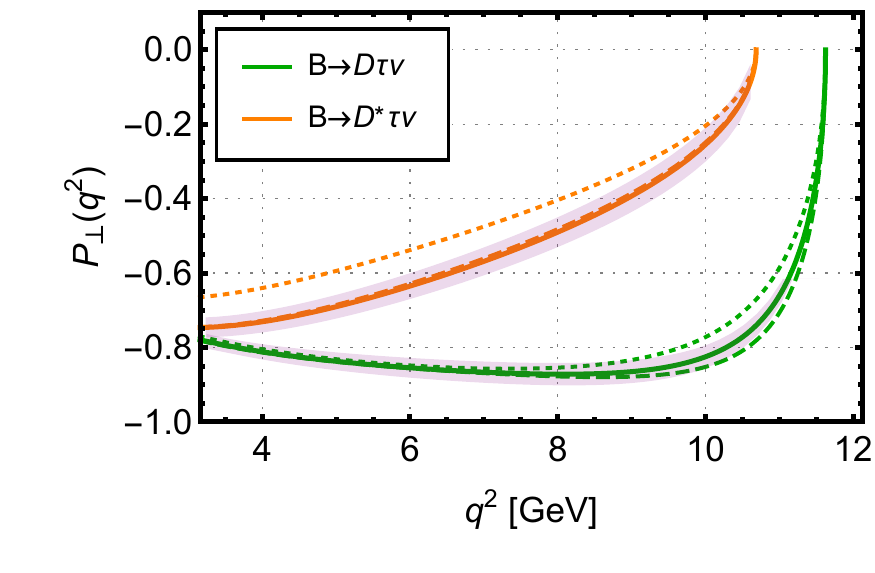} \\
\includegraphics[scale=0.75]{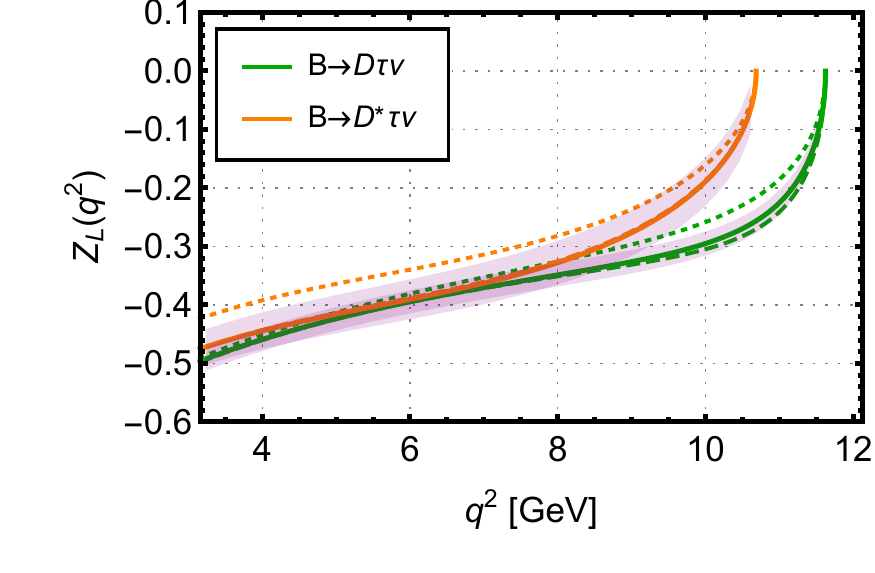}
\includegraphics[scale=0.75]{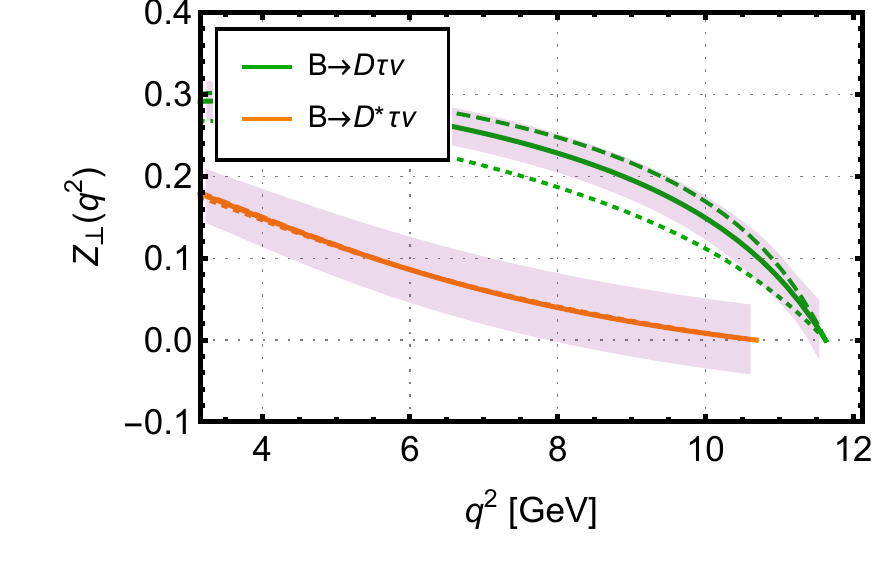} \\
\includegraphics[scale=0.75]{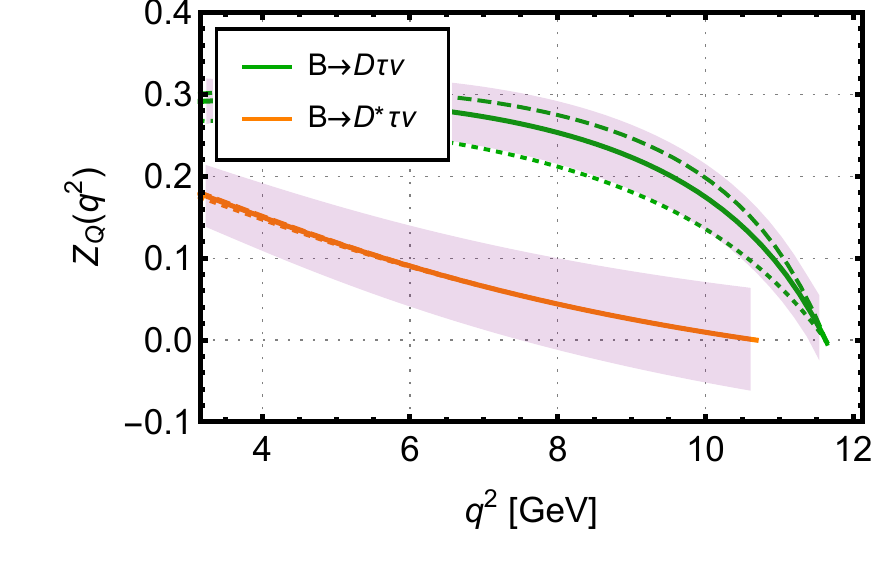} 
\includegraphics[scale=0.75]{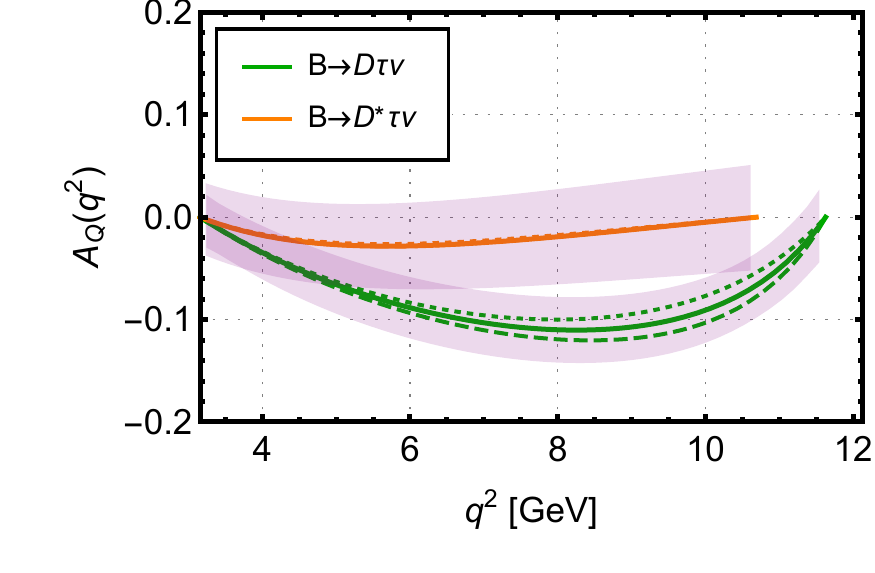} 
\caption{Distributions of $A_{FB}$, $P_L$, $P_{\perp}$, $Z_L$, $Z_\perp$, $Z_Q$ and $A_Q$ for the decays $B\rightarrow D \tau \nu$ (green) and $B\rightarrow D^\ast \tau \nu$ (orange). The solid curves show the SM predictions. The dashed (dotted) curves refer to two NP scenarios with $U_1$ ($S_1 - R_2$) leptoquarks discussed in Section~\ref{sec:new-physics}. The statistical uncertainties obtained from our maximum likelihood analysis are shown as purple bands for $N=3000$ events in the $\tau\to \pi\nu$ channel.}
\label{fig:DstAllObs}
\end{figure}

Figure \ref{fig:DstAllObs} shows the asymmetries as functions of $q^2$ in the SM for $B\to D\tau\nu$ and $B\to D^\ast\tau\nu$, using hadronic form factors of~\cite{Alonso:2016gym}. 
In addition to the SM predictions, we also show the predictions from two benchmark NP scenarios, which are discussed in detail in Section~\ref{sec:new-physics}.

\begin{figure}[phbt]
\centering
\includegraphics[scale=0.35]{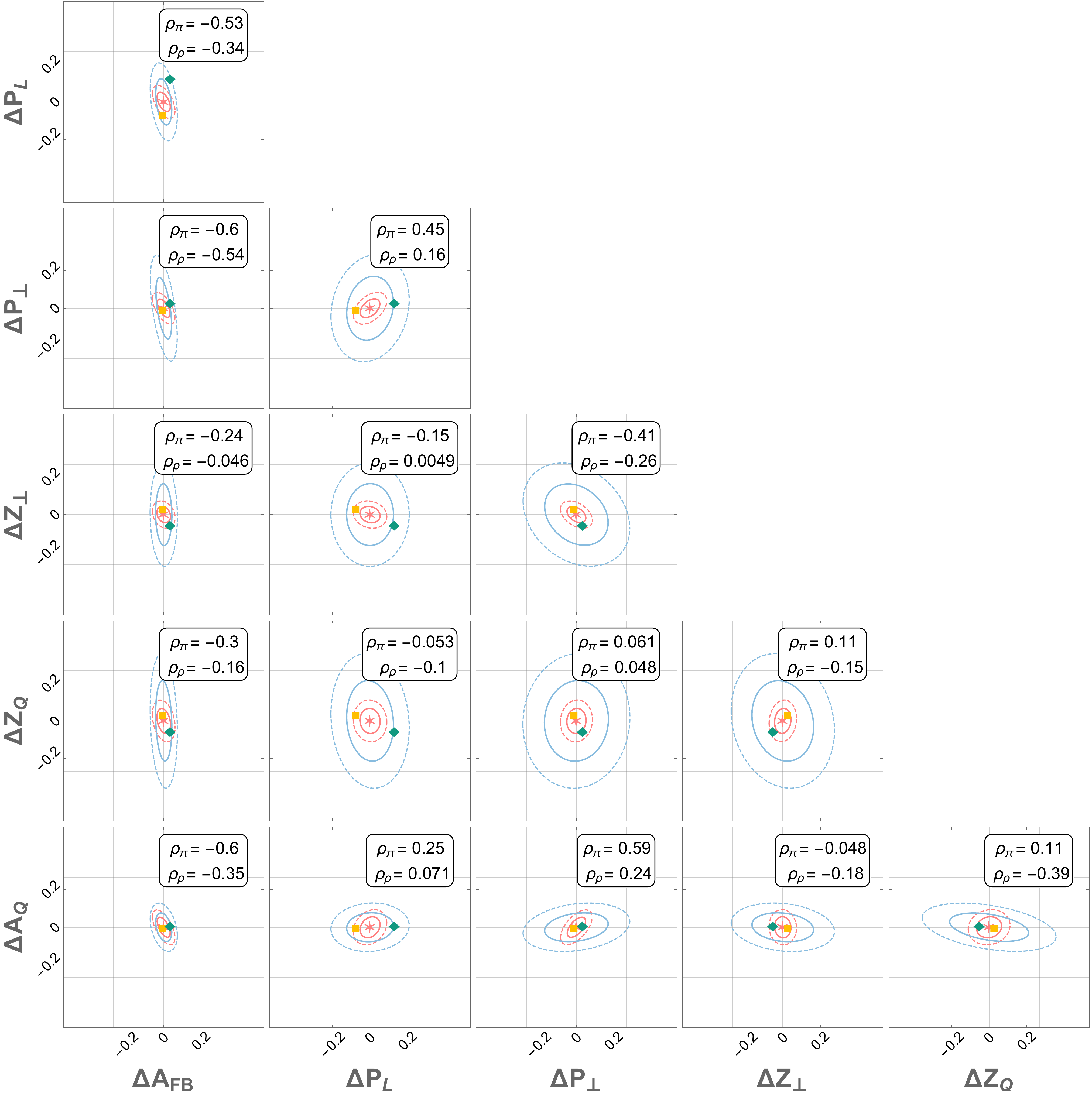}
\caption{$68\%$ (solid) and $95\%$ (dashed) confidence intervals  for the statistical sensitivity to the $\tau$ asymmetries in the SM in $B\rightarrow D \tau(\rightarrow \pi \nu) \nu$ (pink) and $B\rightarrow D \tau(\rightarrow \rho \nu) \nu$ (blue) decays. The central values are marked for the SM (pink star), as well as for the NP scenarios $U_1$ (yellow square) and $S_1-R_2$ (green diamond). The correlation coefficients $\rho_\pi$ and $\rho_\rho$ for each pair of asymmetries are shown in a boxed insert. Assumed is a data set of $N=3000$ total events in each channel.}
\label{fig:gridDSMplusNP}
\end{figure}

\begin{figure}[phbt]
\centering
\includegraphics[scale=0.3]{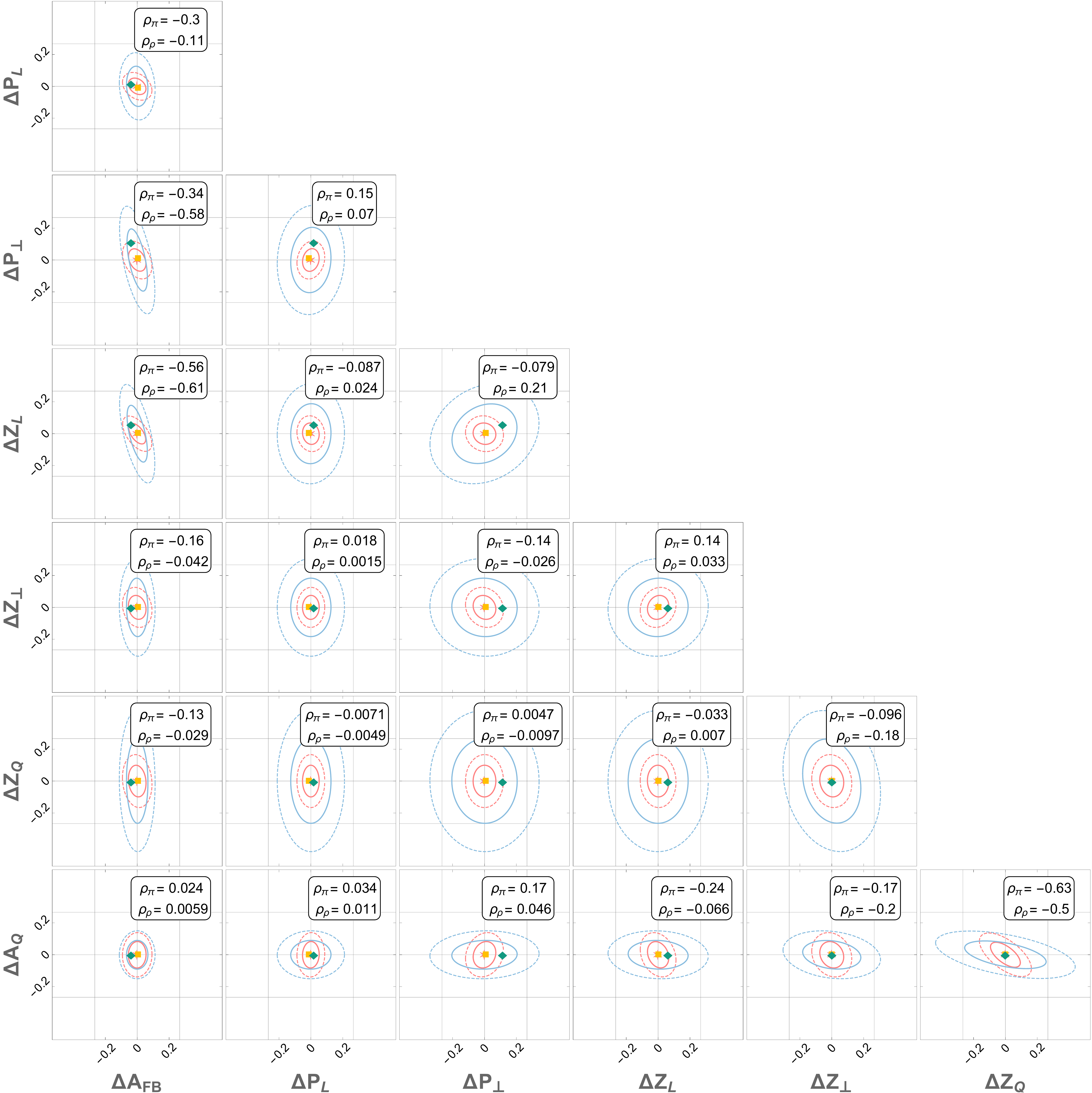}
\caption{$68\%$ (solid) and $95\%$ (dashed) confidence intervals for the statistical sensitivity to the $\tau$ asymmetries in the SM in $B\rightarrow D^\ast \tau(\rightarrow \pi \nu) \nu$ (pink) and $B\rightarrow D^\ast \tau(\rightarrow \rho \nu) \nu$ (blue) decays. The central values are marked for the SM (pink star), as well as for the NP scenarios $U_1$ (yellow square) and $S_1-R_2$ (green diamond). The correlation coefficients $\rho_\pi$ and $\rho_\rho$ for each pair of asymmetries are shown in a boxed insert. Assumed is a data set of $N=3000$ total events in each channel.}
\label{fig:gridDstarSMplusNP}
\end{figure}

The values of the integrated asymmetries in the SM are displayed in Tab.~\ref{tab:sigmas}, along with the projected statistical sensitivities for $N=3000$ total events in each final state $\pi$ or $\rho$.\footnote{The number $N=3000$ is meant to be a {\it very} rough estimate of the number of $B\to D^{(*)}\tau(\to d\nu)\nu$ events expected with 50~ab$^{-1}$ of Belle II data~\cite{Aushev:2010bq,Alonso:2017ktd}.   } Figs.~\ref{fig:gridDSMplusNP} and \ref{fig:gridDstarSMplusNP} visualize these numbers and provide the correlation between each asymmetry pair. The achievable sensitivities for these observables are at the percent level, comparable to the projected sensitivity to $R_{D^{(*)}}$ \cite{belle2}. 

Interestingly, Tab.~\ref{tab:sigmas} suggests that the new observables $Z_L$, $Z_\perp$, $Z_Q$, $A_Q$ could be measured with comparable precision to the previously studied observables $P_L$, $P_\perp$ and $A_{\rm FB}$. We also find that the sensitivities to the asymmetries are comparable in both $B\rightarrow D \tau\nu$ and $B\rightarrow D^* \tau\nu$ decays.  However, there is a stark difference between $\pi$ and $\rho$: for all the observables, the $\tau \rightarrow \pi \nu$ channel has a better sensitivity compared to the $\tau \rightarrow \rho \nu$ channel. Measuring the $\rho$ polarizations would presumably enhance the sensitivity in the latter channel.

Of all the asymmetries in Tab.~\ref{tab:sigmas}, only $P_L$ for $B\rightarrow D^* \tau \nu$ has been measured so far. The projected statistical uncertainty (obtained by rescaling the current measurement with luminosity) is $\pm 0.06$, see Tab.$~50$ in \cite{belle2}; this is in the same ballpark as our projection in Tab.~\ref{tab:sigmas}. The difference may be attributable in part to the background effects we have neglected, as well as detector acceptance and efficiency. Nonetheless, the fact that our purely theoretical estimate of the sensitivity is within a factor of 2 of the official projection provides some confidence in the sensitivity estimates for the other observables.

\begin{table}[h]
\centering
\begin{tabular}{|c|c|cccc|c|}
\hline
\multicolumn{2}{|c|}{}& SM & $\sigma_{\rm th}$ & $\sigma_\pi$ & $\sigma_\rho$ & measured\\
\hline
\multirow{6}{*}{$B\to D\tau\nu$}
& $A_{\rm FB}$ & $-0.359$ & 0.003 & 0.020 & 0.024 & --\\
& $P_L$        & 0.34     & 0.03  & 0.029 & 0.069 & --\\
& $P_\perp$    & $-0.839$ & 0.007 & 0.028 & 0.094 & --\\
& $Z_\perp$    & 0.224    & 0.012 & 0.024 & 0.091 & --\\
& $Z_Q$        & 0.243    & 0.012 & 0.037 & 0.118 & --\\
& $A_Q$        & $-0.088$ & 0.004 & 0.031 & 0.042 & --\\
\hline
\multirow{7}{*}{$B\to D^*\tau\nu$} 
& $A_{\rm FB}$ & 0.07      & 0.02   & 0.031 & 0.037 &--\\
& $P_L$        & $-0.50$   & 0.02   & 0.029 & 0.070 & $-0.38(54)$~\cite{Abdesselam:2016xqt,Hirose:2017dxl}\\
& $P_\perp$    & $-0.49$   & 0.02   & 0.039 & 0.113 & --\\
& $Z_L$        & $-0.323$  & 0.007  & 0.037 & 0.104 & --\\
& $Z_\perp$    & 0.054     & 0.002  & 0.041 & 0.101 &--\\
& $Z_Q$        & $0.058$   & 0.002  & 0.055 & 0.046 & --\\
& $A_Q$        & $-0.0189$ & 0.0005 & 0.146 & 0.050 &--\\
\hline
\end{tabular}
\caption{Numerical predictions of the integrated observables in the SM, together with their theoretical uncertainties $\sigma_{\rm th}$ and the estimated statistical uncertainties in the $\pi$ and $\rho$ channels, $\sigma_\pi$ and $\sigma_\rho$. The theoretical uncertainties are obtained scanning theoretical inputs as in~\cite{Alonso:2016gym}. The statistical uncertainties assume a data set of $N=3000$ events for each final state. Both the theoretical and statistical uncertainties refer to the $68\%$ confidence level.}
\label{tab:sigmas}
\end{table}

\clearpage

\clearpage

\subsection{New physics in $b\to c\tau\nu$}
\label{sec:new-physics}

Heavy new physics at scales $\Lambda \gg m_W$ can modify the total rates and kinematic distributions of the $\tau$ lepton and the $D^{(\ast)}$ meson in the $B\to D^{(\ast)}\tau\nu$ decays. Such modifications can be parameterized in a model-independent way in terms of an effective Hamiltonian
\begin{align}\label{eq:Heff}
\mathcal{H}_{\rm eff} = \frac{4 G_F V_{cb}}{\sqrt{2}}\left(O_{LL}^V + \sum_{X,Y = L,R}\left(C_{XY}^S O_{XY}^S + C_{XY}^V O_{XY}^V\right) + \sum_{X = L,R} C_{XX}^T O_{XX}^T\right),
\end{align}
where $G_F=1/\sqrt{2}v^2$ and $V_{cb}$ is the CKM element. The various effective operators describe local scalar, vector and tensor four-fermion interactions, defined as
\bea
O_{XY}^S & = (\bar{c}\, P_X b)(\bar{\tau}\, P_Y \nu)\,\cr
O_{XY}^V & = (\bar{c}\, \gamma^\mu P_X b)(\bar{\tau}\,\gamma_\mu P_Y \nu)\,\cr
O_{XX}^T & = (\bar{c}\, \sigma^{\mu\nu} P_X b)(\bar{\tau} \,\sigma_{\mu\nu} P_X \nu)\,.
\eea
The Wilson coefficients $C_{XY}^i$ in (\ref{eq:Heff}) contain information pertaining the short-distance structure of the $b\to c\tau\nu$ transitions induced by new physics above the weak scale. In our conventions the SM corresponds to $C^{i}_{XY}=0$. A given NP model induces specific modifications of the Wilson coefficients that can be analyzed by measuring various observables in these decays. We neglect corrections of $\mathcal{O}(v/\Lambda)$ that arise from higher-dimensional operators in the effective theory.

As we discussed in the introduction, current measurements of total rates in terms of the ratios $R_{D^{(*)}}$ are in tension with the SM at a significance of about $3~\sigma$, which could be due to the presence of new physics in $b\to c\tau\nu$ transitions. Several models have been proposed that can explain this difference~\cite{Sakaki:2013bfa,Freytsis:2015qca,Murgui:2019czp,Cheung:2020sbq,Alonso:2015sja,Barbieri:2015yvd,Fajfer:2015ycq,Barbieri:2016las,Assad:2017iib,Calibbi:2017qbu,DiLuzio:2017vat,Bordone:2017bld,Barbieri:2017tuq,Altmannshofer:2017poe,Greljo:2018tuh,Blanke:2018sro,Bordone:2018nbg,Angelescu:2018tyl,Crivellin:2018yvo,Hati:2019ufv,Fuentes-Martin:2020bnh,Li:2016vvp,Crivellin:2017zlb,Cai:2017wry,Marzocca:2018wcf,Aydemir:2018cbb,Becirevic:2018afm,Greljo:2015mma,Boucenna:2016wpr,Megias:2017ove,He:2017bft,Matsuzaki:2017bpp,Iguro:2018fni,Celis:2012dk,Crivellin:2012ye,Crivellin:2015hha,Celis:2016azn,Iguro:2017ysu,Fraser:2018aqj,Greljo:2018ogz,Asadi:2018wea,Robinson:2018gza,Azatov:2018kzb,Mandal:2020htr}.
 One class of models particularly favored by data involve a vector leptoquark $U_1$ transforming as $(\mathbf{3},\mathbf{1},+2/3)$ under the SM gauge group $SU(3)\times SU(2)\times U(1)$. The exchange of such a leptoquark induces $b\to c\tau\nu$ transitions at tree level and generates the operators $O_{LL}^V$ and $O_{RL}^S$~\cite{Alonso:2015sja,Barbieri:2015yvd,Fajfer:2015ycq,Barbieri:2016las,Assad:2017iib,Calibbi:2017qbu,DiLuzio:2017vat,Bordone:2017bld,Barbieri:2017tuq,Altmannshofer:2017poe,Greljo:2018tuh,Blanke:2018sro,Bordone:2018nbg,Angelescu:2018tyl,Crivellin:2018yvo,Hati:2019ufv,Fuentes-Martin:2020bnh}. Another possibility are the scalar leptoquarks $S_1:\,(\mathbf{\bar 3},\mathbf{1},+1/3$) and $R_2:\,(\mathbf{3},\mathbf{2},+7/6)$ that produce a correlated effect in the scalar and tensor operators $O_{LL}^S$ and $O_{LL}^T$~\cite{Sakaki:2013bfa,Li:2016vvp,Crivellin:2017zlb,Cai:2017wry,Marzocca:2018wcf,Aydemir:2018cbb,Becirevic:2018afm}.\footnote{Other models involving colorless gauge bosons $W^\prime$~\cite{Greljo:2015mma,Boucenna:2016wpr,Megias:2017ove,He:2017bft,Matsuzaki:2017bpp,Iguro:2018fni}, extending the Higgs sector~\cite{Celis:2012dk,Crivellin:2012ye,Crivellin:2015hha,Celis:2016azn,Iguro:2017ysu,Fraser:2018aqj} or adding right-handed neutrinos~\cite{Greljo:2018ogz,Asadi:2018wea,Robinson:2018gza,Azatov:2018kzb,Mandal:2020htr} could explain the discrepancy, but they are more in tension with other low-energy observables or collider searches~\cite{Alonso:2016oyd,Akeroyd:2017mhr,Greljo:2018tzh}.  }

We use these two models to demonstrate the sensitivity of our asymmetries to new physics. Our benchmarks correspond to
\begin{align}
\text{``$U_1$ vector leptoquark''}:~& C^V_{LL}=0.08,\quad C^S_{RL}=-0.05,\\\nonumber
\text{``$S_1 - R_2$ scalar leptoquarks''}:~& C^S_{LL}=0.07,\quad C^T_{LL}=-0.03,
\end{align}
where the Wilson coefficients are evaluated at the bottom quark mass scale. Both benchmarks are motivated by a fit to the current $R_D$ and $R_{D^{*}}$ measurements~\cite{Shi:2019gxi}.

\begin{figure}
\centering
\includegraphics[scale=0.3]{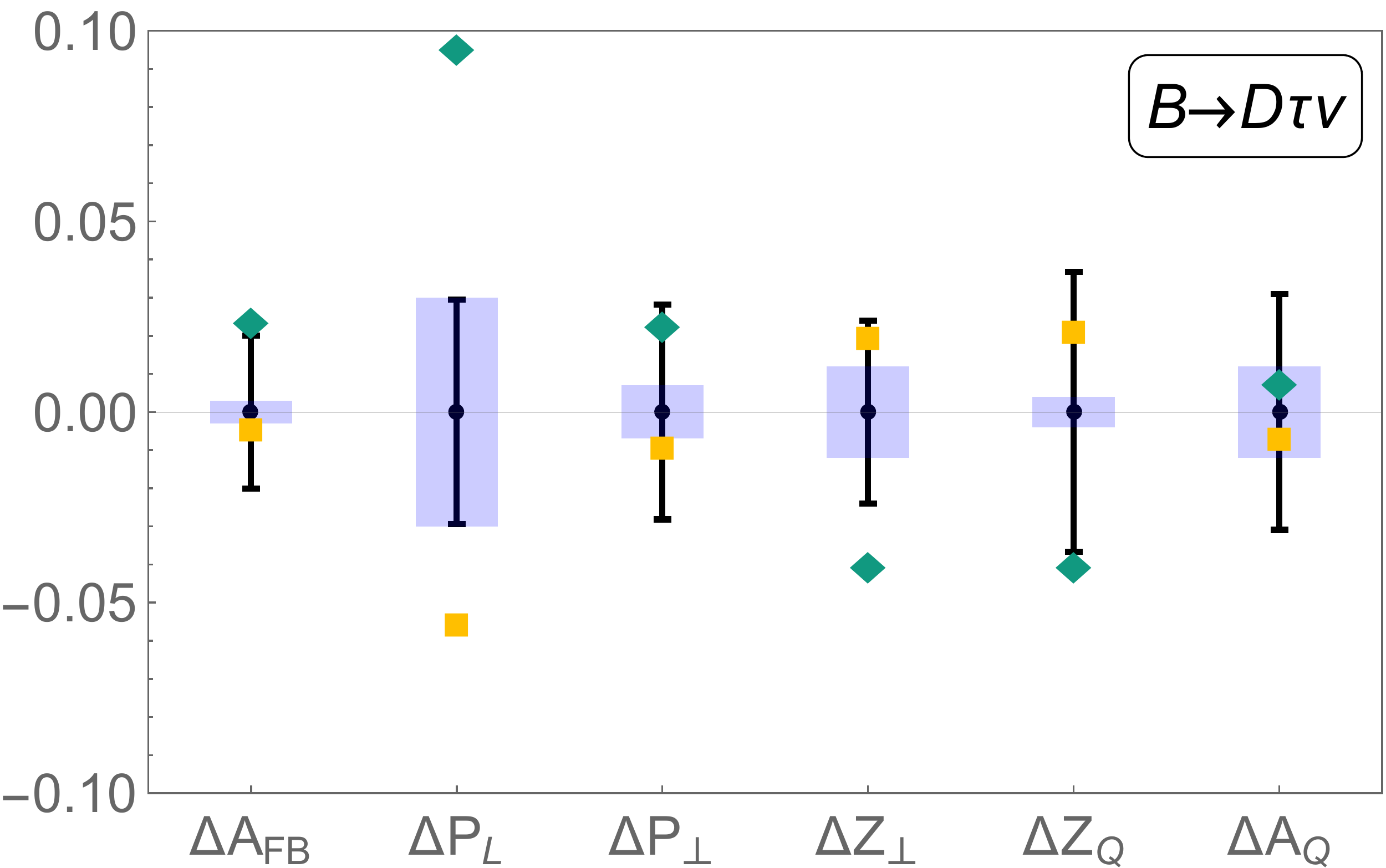}\quad
\includegraphics[scale=0.3]{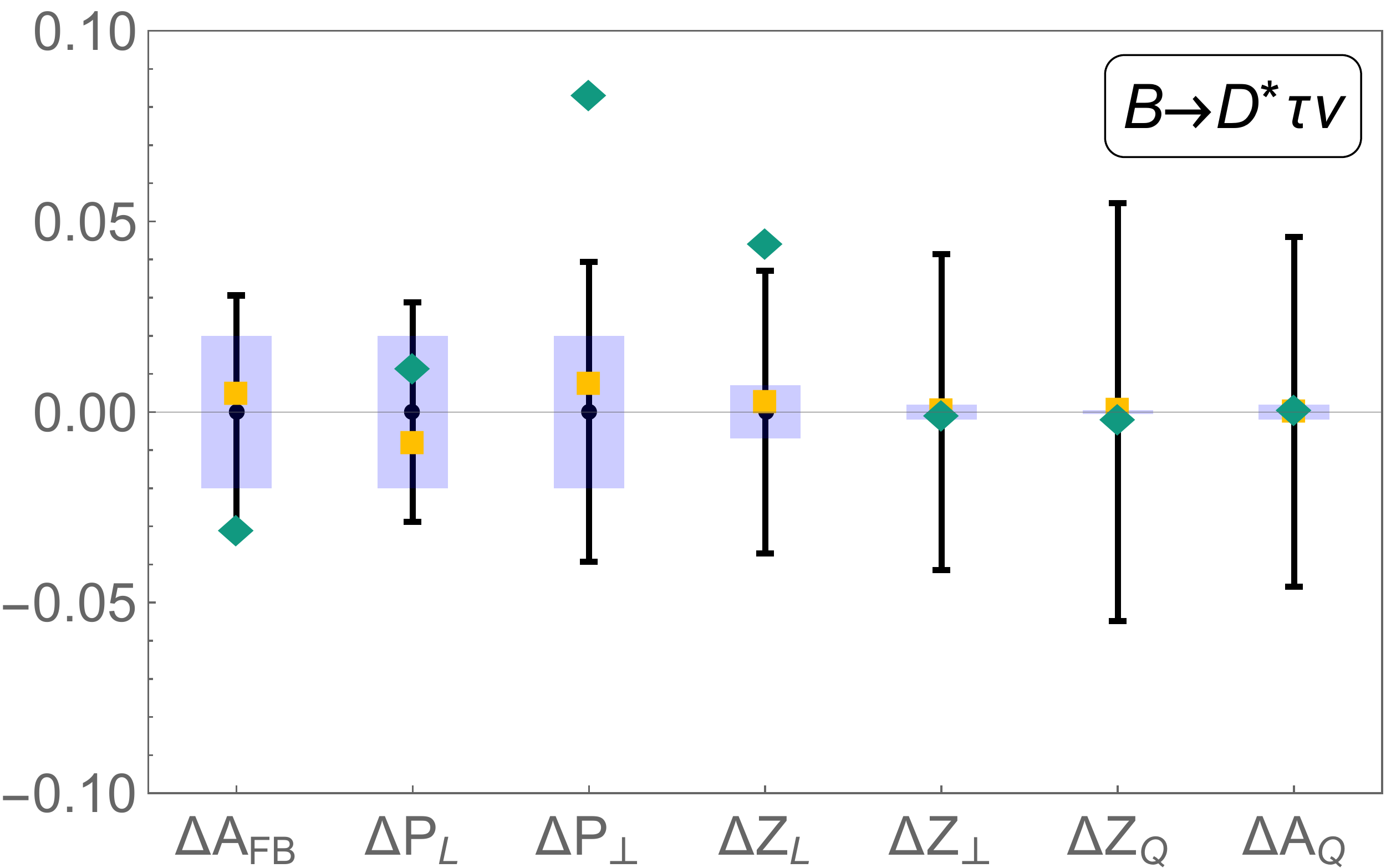}
\caption{Theoretical (blue bands) and statistical (black bars) uncertainties on the asymmetries at 68$\%$ confidence level for $B\rightarrow D\tau\nu$ (left) and $B\rightarrow D^* \tau \nu$ (right). The statistical uncertainties correspond to the $\tau \to \pi\nu$ channel, assuming $3000$ events. The deviations of the central values for the NP scenarios $U_1$ (yellow square) and $S_1-R_2$ (green diamond) from to the SM central values are also shown. }
\label{fig:errorcomp}
\end{figure}

In Fig.~\ref{fig:DstAllObs} we show the $q^2$-dependence of all the asymmetry observables in the $U_1$ (dashed) and the $S_1 - R_2$ (dotted) leptoquark scenarios. We also show the predictions of the $q^2$-inclusive observables in these models in Figs.~\ref{fig:gridDSMplusNP} and \ref{fig:gridDstarSMplusNP} as yellow squares and green diamonds, respectively. We have not included the expected confidence regions around the NP points, but checked that the statistical sensitivities are nearly indistinguishable in size and shape from the SM ellipses. Finally, in Fig.~\ref{fig:errorcomp} we show the $q^2$-integrated results of the observables including both the theoretical and statistical uncertainties for the SM predictions~\cite{Alonso:2016gym}.\footnote{ Fig.~\ref{fig:errorcomp} indicates that the theoretical uncertainties on the observables are always comparable to or smaller than the statistical uncertainties. This further motivates performing the measurement at Belle II, as the precision will not be theoretically limited.  
}

As can be seen from the figures, in the $U_1$ leptoquark scenario most of the observables are very similar to the SM. Only $P_L$ in $B \rightarrow D \tau \nu$ causes an appreciable deviation from the SM prediction. The reason is that the vector leptoquark primarily induces
 the operator $O^V_{LL}$. This effect merely changes the overall normalization of the decay rate in the SM and cancels out in the normalized asymmetry observables. Any observable effect of the $U_1$ leptoquark is due to the small scalar contribution to $O_{RL}^S$, which $P_L$ is especially sensitive to. On the other hand, in the scalar leptoquark scenario the deviation from the SM is quite significant for many of the observables. This scenario involves a combination of scalar and tensor operators, which significantly affect the angular distributions in the $b\to c\tau\nu$ decay.

All in all, we conclude from Fig.~\ref{fig:errorcomp} that the most promising single observables for distinguishing between these two NP scenarios are $P_L(D)$, $P_\perp(D^*)$ and $Z_L(D^*)$. At the same time, 
 no single observable presents a ``slam dunk" case for one NP scenario or the other; differences are at 1-2$\,\sigma$ at best. However, Fig.~\ref{fig:errorcomp} and Figs.~\ref{fig:gridDSMplusNP}-\ref{fig:gridDstarSMplusNP} indicate that the combination of multiple observables offers a way to achieve higher sensitivity.  
This emphasizes the potential of a simultaneous measurement of all of these observables to clarify the nature of the currently observed discrepancies. 

\section{Discussion and conclusion}
\label{sec:discuss}
The study of $B\to D^{(\ast)}\tau\nu$ transitions offers a unique window into couplings between quarks and leptons involving the third generation. 
In this work, we have shown how to extract the maximum information about the
 $b\to c\tau\nu$ transition from kinematic distributions of the observable particles in $B\to D^{(\ast)}\tau(\to \pi \nu,\rho \nu)\nu$ decays. The physics of $B\to D^{(\ast)}\tau\nu$ decays with polarized $\tau$ leptons and unpolarized $D^{(\ast)}$ mesons beyond total rates can be fully captured by nine coefficient functions in a partial wave expansion.
 Linear combinations of some of these functions correspond to widely studied $\tau$ obser\-vables, such as the longitudinal polarization asymmetry $P_L$ and the forward-backward asymmetry $A_{FB}$.
 We showed that seven of the nine coefficient functions can be recovered from the kinematic distributions of the observable particles $D^{(\ast)}$ and $\pi,\rho$. The remaining two functions are sensitive to CP violation and can only be extracted by including additional information, for instance from the $D^{(\ast)}$ decay~\cite{Bhattacharya:2020lfm}. We leave such a study for future work~\cite{future}. 

A similar analysis has previously been performed for a subset of the asymmetries in $B\rightarrow D\tau\nu$~\cite{Alonso:2017ktd}. In this work we generalized this analysis to include $B\rightarrow D^{(\ast)}\tau\nu$ and developed a common framework to describe both processes. Using 
 this framework, we discovered four new asymmetries, $Z_L$, $Z_\perp$, $Z_Q$ and $A_Q$. These observables probe independent fundamental properties of $b\to c\tau\nu$ transitions and can also be extracted from the observable kinematic distributions, which previously had not been realized.

To assess the potential of the Belle II experiment to measure the seven asymmetries, we have performed a statistical analysis assuming the full dataset of 50 ab$^{-1}$. Our unbinned maximum likelihood fit to the fully-differential final-state distribution in $B\to D^{(\ast)}\tau(\to \pi \nu,\rho \nu)\nu$ decays shows that almost all asymmetries could be accessed with a statistical uncertainty of a few percent. These predictions do not include 
realistic experimental effects such as detector acceptance/efficiency/smearing, backgrounds, and systematic uncertainties, see e.g.~\cite{Bernlochner:2020tfi} for further discussion. It would be interesting to further our study by taking these issues into account. 

Additional sensitivity can be obtained by combining the $\tau\to \pi\nu$ and $\tau\to\rho\nu$ channels with each other and with the leptonic decay modes $\tau\to \ell\nu\nu$. While the lepton kinematics do not contain as much information about the asymmetries as the $\pi$ or $\rho$, the leptonic decays occur at a higher rate and should be included in a global analysis of all $\tau$ decay modes. 

These positive measurement prospects and the precise predictions of the asymmetries in the SM allow us to detect possible deviations in the presence of heavy new physics. For two new physics scenarios with scalar and vector leptoquarks, motivated by the currently observed deviations in $B\to D^{(\ast)}\tau\nu$ decays, we have determined the statistical sensitivity compared to the SM expectations. In $B\rightarrow D\tau\nu$, the longitudinal $\tau$ polarization asymmetry $P_L$ discriminates particularly well between the two NP models; in $B\rightarrow D^*\tau\nu$ the perpendicular polarization asymmetry $P_\perp$ and the double asymmetry $Z_L$ show the best individual discriminating power. Of course, the ability to discriminate between different NP models increases by combining all seven asymmetries in a global fit.

In this paper we have endeavored to demonstrate the usefulness of the asymmetries in $B\to D^{(*)}\tau\nu$ decays and the feasibility of measuring them at Belle II. The asymmetries furnish an important intermediate step between the raw data and the underlying Lagrangian parameters, e.g.\ the Wilson coefficients. The framework developed in this paper provides us with a solid interpretation scheme for $\tau$ polarimetry in $B\to D^{(\ast)}\tau\nu$ decays, ready to be confronted with fresh data at Belle II.

\section*{Acknowledgments}

We thank Daniel Aloni, Rodrigo Alonso, Zoltan Ligeti, and Dean Robinson for their constructive comments on the manuscript. The work of PA is supported by DOE grant DE-SC0012567 and MIT Department of Physics. JMC acknowledges support from the Spanish MINECO through the ``Ram\'on y Cajal'' program RYC-2016-20672 and the grant  PGC2018-102016-A-I00. AH and DS are supported by DOE grant DOE-SC0010008. AH and DS are grateful to LBNL, BCTP and BCCP for their generous support and hospitality during the latter's sabbatical year. SW acknowledges support by the German Research Foundation (DFG) under grant no. 396021762–TRR 257.

\appendix

\section{Angular integrals}
\label{sec:angint}

In this appendix we give details about the integration over angles that are not observable in the final state. Using the angular expansions of \eqref{cosexp} and \eqref{tauangles}, the angular integrals in \eqref{d4GammaTod3Gamma} are found to be of the form
\begin{equation}
\label{u-function}
	u(\cos\theta_d)=\int_{-1}^1 d\cos\theta_{\tau} \int_{-\pi}^{\pi} d\zeta \, f(\zeta) \,g(\cos\theta_{\tau})\,\delta(\cos\theta_{\tau}\cos\varphi+\sin\theta_{\tau}\sin\varphi\cos\zeta-\cos\theta_d)\,.
\end{equation}
If $f$ is an odd function of $\zeta$, the integral vanishes. This is the reason why $P_T$ and $Z_T$, which are proportional to $\sin\zeta$ in the total decay rate, vanish. Changing variables from $\zeta$ to $\cos\zeta$, it is straightforward to calculate
\begin{equation}
\label{rhointf1}
\begin{split}
	2\!\int_{-\pi}^{0}d\zeta\, &f(\zeta)\, \delta(\cos\theta_{\tau}\cos\varphi-\sin\theta_{\tau}\sin\varphi\cos\zeta-\cos\theta_d)\\
	&=2\int_{-1}^1 d\cos\zeta\, f(\cos\zeta) \,\frac{1}{\sqrt{1-\cos^2\zeta}}\frac{1}{|\sin\theta_{\tau}\sin\varphi|}\delta\left(\cos\zeta-\cos\zeta_0\right)\\
	&=2\, |\mathrm{det}\,\mathbf{J}|f(\cos\zeta_0)\,,
\end{split}
\end{equation}
where
\begin{equation}
	\cos\zeta_0=\frac{\cos\theta_d-\cos\theta_{\tau}\cos\varphi}{\sin\theta_{\tau}\sin\varphi}
\end{equation}
and the Jacobian is given by
\begin{equation}
	|\mathrm{det}\mathbf{J}|=\big(1-\cos^2\theta_d-\cos^2\theta_{\tau}-\cos^2\varphi+2\cos\theta_d\cos\theta_{\tau}\cos\varphi\big)^{-1/2}\,.
\end{equation}
After integrating over $\zeta$, the delta function in \eqref{rhointf1} restricts the possible range of $\theta_{\tau}$. Solving \eqref{changeofvariables} for $\cos\theta_{\tau}$ and inserting the $\zeta$ integration limits gives 
\begin{equation}
	\cos\theta_{\tau}\big|_{\zeta=0}=\cos(\theta_d \mp \varphi)\,, \quad \cos\theta_{\tau}\big|_{\zeta=\pm\pi}=\cos(\theta_d \pm \varphi)\,.
\end{equation}	
The choice of sign configuration does not matter, since the other configuration can be obtained by sending $\varphi\rightarrow -\varphi$; this angle is only defined in terms of $\cos\varphi$.
 Choosing $\varphi\geq 0$ and $\cos(\theta_d + \varphi)$ as the lower integration limit gives 
\begin{equation}
\begin{split}
		u(\cos\theta_d)&=\int_{-1}^1 \,d\cos\theta_{\tau} \, g(\cos\theta_{\tau}) \, \int_{-\pi}^{\pi} d\zeta\,f(\zeta)\,  \delta(\cos\theta_{\tau}\cos\varphi-\sin\theta_{\tau}\sin\varphi\cos\zeta-\cos\theta_{d})\\
	&=2  \int_{\cos(\theta_d + \varphi)}^{\cos(\theta_d - \varphi)} d\cos\theta_{\tau} \, |\mathrm{det}\,\mathbf{J}| f(\cos\zeta_0)\,g(\cos\theta_{\tau})\,.
\end{split}
\end{equation}
The procedure above is equivalent to the change of variables in~\cite{Alonso:2016gym}. The resulting functions $u(\cos\theta_d)$ for all functions $f(\zeta)g(\cos\theta_{\tau})$ present in the full decay rate are listed in Tab.~\ref{tab:angularintegral}.
\begin{table}[!t]
\centering
\begin{tabular}{|c|c|}
\hline 
$f(\zeta)g(\cos\theta_{\tau})$ & $u(\cos\theta_d)$ \\ 
\hline 
1 & $2\pi$ \\ 
\hline 
$\cos\theta_{\tau}$ & $2\pi\cos\theta_d\cos\varphi$ \\ 
\hline 
$\cos^2\theta_{\tau}$ & $2\pi(\cos^2\theta_d\cos^2\varphi+\frac{1}{2}\sin^2\theta_d \sin^2\varphi)$ \\ 
\hline 
$\cos\zeta\sin\theta_{\tau}$ & $2\pi\cos\theta_d\sin\varphi$ \\ 
\hline 
$\cos\zeta\sin(2\theta_{\tau})/2$ & $\pi\sin\varphi\cos\varphi(3\cos^2\theta_d-1)$ \\ 
\hline 
$\sin\zeta\sin(\theta_{\tau})$ & $0$ \\ 
\hline 
$\sin\zeta\sin(2\theta_{\tau})/2$ & $0$ \\ 
\hline 
\end{tabular} 
\caption{Angular integrals from equation \eqref{u-function}.}
\label{tab:angularintegral}
\end{table}

\section{Analytical expressions for the new asymmetries}
\label{appx:analytic}
In this appendix we report the analytic expressions for all new asymmetries introduced in Section~\ref{sec:obsB}, i.e., $Z_L$, $Z_\perp$, $Z_T$, $A_Q$ and $Z_Q$. Similar formulas for the remaining asymmetries can be found in the appendix of~\cite{Asadi:2018sym}.\footnote{What we call $\frac{d\Gamma}{dq^2} \mathcal{O} (q^2)$ in this work, corresponds to $\frac{d\mathcal{O}}{dq^2}$ in~\cite{Asadi:2018sym}. }

In $B\rightarrow D \tau \nu$, the asymmetries are
\begin{align}
\label{eq:AsD}
Z_L (q^2) & = A_{FB} (q^2)\,, \\\nonumber
\frac{d\Gamma}{dq^2} Z_\perp (q^2) & = \mathcal{N} (m_D,q^2) \mathrm{Re} \left[ \Xi  \right]\,,\\	\nonumber 
\frac{d\Gamma}{dq^2} Z_T (q^2) & = -\mathcal{N} (m_D,q^2)  \mathrm{Im} \left[ \Xi  \right]\, ,	 
\end{align}
with
\begin{align}
\label{eq:SigmaD}
\Xi &=\left(			\left(1+\CVLL+\CVRL\right) \sqrt{q^2} H_{V,0}^s - 4 \CTLL  m_\tau H_T^s		\right) 	\nonumber \\
 &  \quad \times\left(			\left(1+\CVLL+\CVRL\right)^{*} m_\tau H_{V,0}^s - 4 \CTLL^*  \sqrt{q^2} H_T^s		\right)\,,  \\
\mathcal{N} (m_D,q^2)  &= \frac{G_F^2 V_{cb}^2}{192 m_B^3 \pi^3}   \sqrt{\left(	(m_B-m_D)^2-q^2 \right) \left(	(m_B+m_D)^2-q^2 \right)  }  \left(1-\frac{m_\tau^2}{q^2}\right)^2 \,, \nonumber
\end{align}
where $C^X_{YZ}$ refer to the Wilson coefficients of the relevant dimension-6 operators, see \eqref{eq:Heff}. The hadronic functions $H$ can be found in Refs.~\cite{Sakaki:2013bfa,Asadi:2018sym}.

The quadrupole observables for the same decay are given as 
\begin{align}
\label{eq:A2D}
\frac{d\Gamma}{dq^2} A_Q (q^2) &=  \frac{\mathcal{N} (m_D,q^2)}{2 }  \left\lbrace \Big|  	\left(1+\CVLL+\CVRL\right) m_\tau H_{V,0}^s - 4 \CTLL  \sqrt{q^2} H_T^s		\Big|^2  \right.  \\
&\quad - \left. \Big|  		\left(1+\CVLL+\CVRL\right) \sqrt{q^2} H_{V,0}^s - 4 \CTLL  m_\tau H_T^s	\Big|^2 \right\rbrace, \nonumber\\
\frac{d\Gamma}{dq^2} Z_Q (q^2) &=  \frac{\mathcal{N} (m_D,q^2)}{2 }  \left\lbrace \Big|  	\left(1+\CVLL+\CVRL\right) m_\tau H_{V,0}^s - 4 \CTLL  \sqrt{q^2} H_T^s		\Big|^2  \right.  \\
& \quad+ \left. \Big|  		\left(1+\CVLL+\CVRL\right) \sqrt{q^2} H_{V,0}^s - 4 \CTLL  m_\tau H_T^s	\Big|^2 \right\rbrace. \nonumber
\end{align}

Similarly, for $B\rightarrow D^* \tau \nu$ the asymmetries are given by 
\begin{align}
\label{eq:AsDs}
\frac{d\Gamma}{dq^2} Z_L (q^2) &=   \frac{4\mathcal{N} (m_{D^*},q^2)}{3 } \mathrm{Re} \left\lbrace 2 \left(			(\CSRL-\CSLL) \sqrt{q^2} H_{S} + (1+\CVLL-\CVRL) m_\tau H_{V,t}		\right) 	\right. \nonumber  \\
 &\quad\times   \left(		-	(1+\CVLL-\CVRL)^{*} m_\tau H_{V,0} + 4 \CTLL^*  \sqrt{q^2} H_{T,0}		\right)     \\
 & \quad- \left(	(1+\CVLL-\CVRL) \sqrt{q^2} \left(	H_{V,-} + H_{V,+}	\right) + 4 \CTLL m_\tau  \left(	H_{T,-} - H_{T,+}	\right)	\right) \nonumber \\
  &\quad \times  \left. \left(	(1+\CVLL+\CVRL)^* \sqrt{q^2} \left(	H_{V,-} - H_{V,+}	\right) + 4 \CTLL^* m_\tau  \left(	H_{T,-} + H_{T,+}	\right)	\right) \right\rbrace, \nonumber  \\
\frac{d\Gamma}{dq^2} Z_\perp (q^2) & =  - \frac{\mathcal{N} (m_{D^*},q^2)}{2 } \mathrm{Re} \left[  \Xi^* \right], \nonumber   \\
\frac{d\Gamma}{dq^2} Z_T (q^2) &=  \frac{\mathcal{N} (m_{D^*},q^2)}{2 }  \mathrm{Im} \left[  \Xi^* \right] \nonumber  ,  
\end{align}
where 
\begin{align}
\label{eq:Sigma}
\Xi^* &= \left\lbrace  	\left(  \big| 1+ \CVLL \big|^2 + \big|  \CVRL \big|^2  \right) m_\tau \sqrt{q^2} \left(	H_{V,+}^2+	H_{V,-}^2	\right)		 \right. \\
& \quad	-2 \big| 1+\CVLL -\CVRL \big|^2 m_\tau \sqrt{q^2} H_{V,0}^2  + 16 \big| \CTLL \big|^2 m_\tau \sqrt{q^2}  \left(	H_{T,-}^2 + H_{T,+}^2 - 2 H_{T,0}^2	\right)  \nonumber \\
& \quad+ 8 \Big(\left(1+\CVLL-\CVRL\right) \CTLL^* q^2 + \left(1+\CVLL-\CVRL\right)^* \CTLL m_\tau^2\Big) H_{V,0} H_{T,0}   \nonumber \\
& \quad-4 H_{V,+}  \Big( \left(	\CVRL \CTLL^* q^2 + \CVRL^* \CTLL m_\tau^2	\right) H_{T,-}  \nonumber \\
&\quad+ \left(		(1+\CVLL) \CTLL^* q^2 + \CTLL (1+\CVLL)^* m_\tau^2	\right) H_{T,+}  \Big) \nonumber \\
& \quad+2 H_{V,-} \Big(	- 2  \mathrm{Re} \left[ (1+ \CVLL)^* \CVRL \right] m_\tau \sqrt{q^2} H_{V,+} + 2    (1+ \CVLL)^*  \CTLL m_\tau^2 H_{T,-} \nonumber \\
& \quad +  	 2 \CVRL^*\CTLL m_\tau^2 H_{T,+}	+ 2 \left. 	(1+\CVLL)  \CTLL^* q^2 H_{T,-} + 2 \CVRL  \CTLL^* q^2 H_{T,+}  \Big) \right\rbrace , \nonumber
\end{align}
while the quadrupole observables $A_Q$ and $Z_Q$ are
\begin{align}
\label{eq:A2Ds}
\frac{d\Gamma}{dq^2} A_Q (q^2) &= - \frac{\mathcal{N} (m_{D^*},q^2)}{4 } \left\lbrace  \left(	\big|1+\CVLL\big|^2 + \big|\CVRL\big|^2	\right) m_\tau^2  \left( H_{V,-}^2 + H_{V,+}^2 \right)   \right. \nonumber \\
& \quad-2 \big|1+\CVLL-\CVRL\big|^2 m_\tau^2 H_{V,0}^2  \nonumber \\ 
&\quad+ 16 \big|\CTLL\big|^2 q^2 \left( H_{T,-}^2 -2 H_{T,0}^2 + H_{T,+}^2 \right) \nonumber \\
& \quad+ 4m_\tau H_{V,-} \mathrm{Re}\Big[	- (1+\CVLL)\CVRL^* m_\tau H_{V,+}  \nonumber \\
& \quad  + 2\CTLL \sqrt{q^2} \left(		(1+\CVLL)^* H_{T,-} + \CVRL^* H_{T,+}		\right) 	\Big]  \\
& \quad- 8 m_\tau \sqrt{q^2} H_{V,+}  \left( \mathrm{Re} \left[ \CTLL \CVRL^* \right] H_{T,-} + \mathrm{Re} \left[ \CTLL (1+\CVLL)^* \right] H_{T,+}  \right) \nonumber \\
&\quad+ 16  \mathrm{Re} \left[\CTLL \left(	 1+\CVLL-\CVRL	\right)^* \right] m_\tau \sqrt{q^2} H_{V,0} H_{T,0}   \nonumber \\
&\quad -\Big| 			\sqrt{q^2} \left( 	(1+\CVLL) H_{V,-} - \CVRL H_{V,+} 	\right)			+ 4 \CTLL m_\tau H_{T,-}		 \Big|^2   \nonumber \\
& \quad+2\,\Big| 			\sqrt{q^2}  	(1+\CVLL-\CVRL) H_{V,0} - 4 \CTLL m_\tau H_{T,0}	 \Big|^2  \nonumber   \\
&\quad  \left. -\Big| 			\sqrt{q^2} \left(  \CVRL	H_{V,-} - (1+\CVLL)   H_{V,+} 	\right)			+ 4 \CTLL m_\tau H_{T,+}	 \Big|^2   \right\rbrace ,  \nonumber \\ 
\frac{d\Gamma}{dq^2} Z_Q (q^2) &= - \frac{\mathcal{N} (m_{D^*},q^2)}{4 } \left\lbrace  \left(	\big|1+\CVLL\big|^2 + \big|\CVRL\big|^2	\right) m_\tau^2  \left( H_{V,-}^2 + H_{V,+}^2 \right)   \right. \nonumber \\
& \quad-2 \,\big|1+\CVLL-\CVRL\big|^2 m_\tau^2 H_{V,0}^2  \nonumber \\ 
&\quad+ 16 \big|\CTLL\big|^2 q^2 \left( H_{T,-}^2 -2 H_{T,0}^2 + H_{T,+}^2 \right) \nonumber \\
&\quad+ 4m_\tau H_{V,-} \mathrm{Re}\Big[	-(1+\CVLL)\CVRL^* m_\tau H_{V,+} \nonumber \\
& \quad  + 2\CTLL \sqrt{q^2} \left(		(1+\CVLL)^* H_{T,-} + \CVRL^* H_{T,+}		\right) 	\Big]  \\
& \quad- 8 m_\tau \sqrt{q^2} H_{V,+}  \left( \mathrm{Re} \left[ \CTLL \CVRL^* \right] H_{T,-} + \mathrm{Re} \left[ \CTLL (1+\CVLL)^* \right] H_{T,+}  \right) \nonumber \\
&\quad+ 16   \mathrm{Re} \left[\CTLL \left(	 1+\CVLL-\CVRL	\right)^* \right] m_\tau \sqrt{q^2} H_{V,0} H_{T,0}  \nonumber \\
&\quad +\Big| 			\sqrt{q^2} \left( 	(1+\CVLL) H_{V,-} - \CVRL H_{V,+} 	\right)			+ 4 \CTLL m_\tau H_{T,-}		 \Big|^2   \nonumber \\
&\quad -2\Big| 			\sqrt{q^2}  	(1+\CVLL-\CVRL) H_{V,0} - 4 \CTLL m_\tau H_{T,0}	 \Big|^2   \nonumber  \\
&\quad  \left. +\Big| 			\sqrt{q^2} \left(  \CVRL	H_{V,-} - (1+\CVLL)   H_{V,+} 	\right)			+ 4 \CTLL m_\tau H_{T,+}	 \Big|^2   \right\rbrace .  \nonumber
\end{align}
The hadronic functions $H$ are pure functions of $q^2$ and contain the hadronic matrix elements. All theory uncertainties are therefore contained in these functions.

\phantomsection
\addcontentsline{toc}{section}{References}

\bibliographystyle{utphys_modified}
\bibliography{taupolarizations}

\end{document}